\documentclass[a4paper,showpacs,amsmath,amssymb,12pt]{article}

\usepackage{mathrsfs}
\usepackage{amsfonts,graphicx,color,slashbox}
\usepackage{indentfirst}
\usepackage{amssymb}
\usepackage{amsmath}

\numberwithin{equation}{section}
\def \bar {\overline}

\title{D1-D3 (or $\bar{\textrm{D}3}$) Systems with Fluxes }
\author{Bin Chen and Xiao Liu\footnote{Email:bchen01,liuxiaoerty@pku.edu.cn,}\\
{\small Department of Physics} \\
{\small and State Key Laboratory of Nuclear Physics and Technology,}\\
{\small Peking University, Beijing 100871, P.R.China}}
\date{}

\begin{document}

\maketitle

\begin{abstract}
In this article we study D1-D3 (or $\bar{\textrm{D}3}$) brane
systems with generic constant electric and magnetic fluxes in IIB
string theory.  We work out all possible supersymmetric
configurations and find out via T-duality all of them and
corresponding supersymmetry conditions could be related to the
supersymmetric intersecting D1-D1 pairs. And we do D1-D3 (or
$\bar{\textrm{D}3}$) open string quantization  for a class of
configurations. We find that there are many near massless states
in NS sector for near-BPS configurations. Furthermore we calculate
open string pair creation rate in generic nonsupersymmetric
configurations.
\end{abstract}

\newpage

\section{Introduction}

D-branes and anti-D-branes are nonperturbative objects in string
theory, carrying RR charges. A single BPS D-brane preserve half
supersymmetries in flat spacetime. But if we put two different
D-branes together, or let a D-brane be in a compact configuration,
supersymmtry could be broken\cite{book1,book2}. One typical
example is a Dp-anti-Dp system which breaks all the
supersymmetries. However, it turns out that when one turns on the
background flux on the D-branes, the supersymmetries could be
recovered. It is an interesting issue to look for these
configurations.    In \cite{supertube}, Mateos and Townsend found
that if one turned on suitable gauge fluxes on tubular D2-brane,
the system can be supersymmetric. This is so-called `supertube'.
The configuration could be taken as the blow-up of D-particles and
has no net D2 brane charge. Effectively such system may be
simplified to a D2-anti-D2 system with fluxes\cite{tac}. This
discovery led to a lot of study of supersymmetric Dp-(anti)-Dp
system with background fluxes \cite{chen,bran,myers,zoo}.  In
particular, it has been found even in Dp-anti-Dp system, the
system could be supersymmetric if one turns on suitable fluxes.

Another class of non-BPS brane configuration is Dp-Dq  with $p -p
\neq 0$ (mod 4). In particular, D0-D2 system is remarkable. The
straightforward calculation of the D0-D2 open string spectrum
shows that the ground state is tachyonic, which means that the
system is unstable and non-BPS. However, it has been shown in
\cite{Witten96} that  D0-D2 system is actually dual to the (F1,D1)
bound state. In fact, the underlying picture is that through
tachyon condensation D0-D2 system settle down to a D0-D2 bound
state, with D0 being dissolved into D2. Moreover, if one considers
the gauge fluxes on D-brane, the story become more interesting. In
D0-D2 system with magnetic field, in the zero-slop limit, there
could exist an infinite tower of near massless states in the open
string spectrum. On the other hand, the system could be studied in
the framework of noncommutative gauge theory. In \cite{Minwalla},
the authors showed that in D0-D2 system with large magnetic field
the D-particle could be taken as the soliton in (2+1)-dimensional
noncommutative Yang-Mills theory. The large tower of near massless
states corresponds to the fluctuations around the soliton
solution. And due to the existence of the large magnetic field,
the system is actually near-BPS, and the tachyon condensation
could be very well studied, as shown in \cite{Minwalla}. Another
interesting aspect is that there exist BPS configuration in Dp-Dq
($p\neq q$) system if suitable constant background magnetic fluxes
were turned on\cite{ChenB,Witten99}.

It would be interesting to see if there exist BPS configuration in
Dp-(anti)-Dq system with generic background fluxes. As the first
step, in this paper, we will pay attention to the system with
parallel D1-brane and $\textrm{D}3$ (or $\bar{\textrm{D}3}$)-brane
in flat spacetime background. We will turn on constant fluxes on
them, including generic electric and magnetic fluxes. We will try to
find the most general supersymmetric configurations by
 using the  $\Gamma$ matrix method\cite{superd,symmetry}. We get
the necessary conditions that the fluxes must satisfy. In two
simplest setup, we obtain the sufficient condition directly. For
more complicated cases, we try to attack the problem in another way.
We find all possible supersymmetric configurations, which are
related to the two simple cases via T-duality and Lorentz
transformation. We also find that the supersymmetric configurations
are equivalent to the systems studied in \cite{chen}.

Besides looking for BPS configurations, there are other interesting
issues to address in fluxed D1-D3 system. For generic flux setup,
the system is nonsupersymmetric.  The first step to investigate the
nonsupersymmetric configurations is to do quantization of the open
string between D1 and D3 (or $\bar{\textrm{D}3}$). This is a quite
difficult problem due to perplexing boundary conditions imposed at
the ends of the open string. The excitations on the string could
have non-integer (or non-half integer) or even complex modes. There
could be tachyonic excitation and it would be interesting to study
the tachyon condensation in the system. In the D0-Dp system with
constant magnetic fields, it has been found that there could exist
large number of near-massless states if one tune the fluxes
carefully so that the system is near-BPS\cite{seiwit, ChenB}. For
the cases in the paper, we will show that this phenomenon also
happen. Another interesting issue is the open string pair
production\cite{Bachas, pair}. This will happen when the open string
between D1 and D3 (or $\bar{\textrm{D}3}$) have complex modes. We
will calculate the rate of string pair creation from one-loop vacuum
amplitude of 1-3 strings.

This paper is organized as follows. In section 2, we will
introduce the system and work out the supersymmetric
configurations. First, we use $\Gamma$ matrix to discuss
supersymmetry conditions. We find necessary conditions for general
systems and sufficient conditions for some simplified model. After
that, via T-duality and Lorentz transformation, we will finish
supersymmetry discussions and obtain all possible
supersymmetric configurations in fluxed D1-D3
($\bar{\textrm{D}3}$) system. Moreover, we will study the relation
of fluxed D1-D3 system with D-string at angles with relative
motion. In section 3, we will do mode expansion and quantization
of open strings stretched between D1 and D3 (or
$\bar{\textrm{D}3}$). In 3.1, we will calculate open string pair
creation rate when there exist
complex excitation modes. 
In section 3.2, We will determine GSO projection using in section
3.1. And we will study the open string spectrum when the system is
near BPS. In section 4, we will give conclusions and discussions.
In Appendix A, we will present how to get all the necessary
supersymmetry conditions in section 2. In Appendix B and C, we
will give details of T-dual discussions, mode expansions and
quantization.

\section{Supersymmetric configurations}

We would like to study the D1-D3 ($\bar{\textrm{D}3}$)-brane
system in flat spacetime background. Let D1-brane lie along
$X^0,X^1$ and D3 (or $\bar{\textrm{D}3}$)-brane along $X^0,\cdots
X^3$. We will turn on all possible constant fluxes. On D1-brane,
there is only electric flux:
\begin{eqnarray}\label{d1flux}
\tilde{F}_{D1}=\frac{1}{2\pi\alpha'}\left(\begin{array}{cc}
0&-\tilde{E}\\
\tilde{E}&0
\end{array}\right).
\end{eqnarray}
On D3 (or $\bar{\textrm{D}3}$)-brane, we can turn on three
electric fluxes and three magnetic fluxes. But using rotational
symmetry, we can let electric fluxes on D3 (or
$\bar{\textrm{D}3}$) be only in planes $X^0$-$X^1$ and
$X^{0}$-$X^{2}$ without losing generality:
\begin{eqnarray}\label{d3flux}
F_{D3\  ( \bar{D3})}=\frac{1}{2\pi\alpha'}\left(\begin{array}{cccc}
0&-E_{1}&-E_{2}&0\\
E_{1}&0&B_{3}&-B_{2}\\
E_{2}&-B_{3}&0&B_{1}\\
0&B_{2}&-B_{1}&0
\end{array}\right)
\end{eqnarray}

The corresponding DBI action of D1 and D3 are
respectively\footnote{In the following part of this article,
except somewhere in section 3, we will let $2\pi\alpha'=1$.}
\begin{equation}
\mathscr{L}_{1}=\sqrt{-\textrm{det}(g+\tilde{F}_{D1})}=\sqrt{1-\tilde{E}^{2}}
\end{equation}
and
\begin{eqnarray}
\mathscr{L}_{2}&=&\sqrt{-\textrm{det}(g+F_{D3\  (or \bar{D3})})}
\nonumber\\
&=&\sqrt{1-E_{1}^{2}-E_{2}^{2}+B_{1}^{2}+B_{2}^{2}+B_{3}^{2}-(E_{1}B_{1}+E_{2}B_{2})^{2}}
\end{eqnarray}
In this paper, we do not discuss critical cases when
$\mathscr{L}_{1}=0$ or $\mathscr{L}_{2}=0$,  so we require that
\begin{equation}\label{defcon1}
1-\tilde{E}^{2}> 0
\end{equation}
and
\begin{equation}\label{defcon2}
1-E_{1}^{2}-E_{2}^{2}+B_{1}^{2}+B_{2}^{2}+B_{3}^{2}-(E_{1}B_{1}+E_{2}B_{2})^{2}>
0
\end{equation}

According to \cite{superd,symmetry}, the conditions for
supersymmetry is that there exist nonzero $\epsilon$ satisfying both
\begin{eqnarray}\label{basic}
\Gamma^{(1)}\epsilon&=&\epsilon,
\nonumber\\
\Gamma^{(2)}\epsilon&=&\pm\epsilon,
\end{eqnarray}
where $+$ is for D3, $-$ is for $\bar{\textrm{D}3}$, and
$\Gamma^{(1)}$ and $\Gamma^{(2)}$ are the Gamma matrices for D1 and
D3 respectively
\begin{eqnarray}\label{defg1}
\Gamma^{(1)}=\frac{1}{\sqrt{1-\tilde{E}^{2}}}\left(\begin{array}{cc}
0&\Gamma_{01}-\tilde{E}\\
\Gamma_{01}+\tilde{E}&0
\end{array}\right)
\end{eqnarray}
\begin{eqnarray}\label{defg2}
\Gamma^{(2)}&=&\frac{1}{\sqrt{1-E_{1}^{2}-E_{2}^{2}+B_{1}^{2}+B_{2}^{2}+B_{3}^{2}-(E_{1}B_{1}+E_{2}B_{2})^{2}}}\times
\nonumber\\
&&\left[\left(\begin{array}{cc}
0&K\\
K&0
\end{array}\right)+\left(\begin{array}{cc}
0&\Gamma_{0123}-E_{1}B_{1}-E_{2}B_{2}\\
-\Gamma_{0123}+E_{1}B_{1}+E_{2}B_{2}&0
\end{array}\right)
\right]\nonumber\\
\end{eqnarray}
with
\begin{equation}
K=-E_{1}\Gamma_{23}+E_{2}\Gamma_{13}+B_{1}\Gamma_{01}+B_{2}\Gamma_{02}+B_{3}\Gamma_{03}.
\end{equation}

 Because IIB theory is chiral, $\epsilon$  must also satisfy
\begin{eqnarray}
\tilde{\Gamma}_{11}\epsilon=\left(\begin{array}{cc}
\Gamma_{11}&0\\
0&\Gamma_{11}
\end{array}\right)\epsilon=\epsilon.
\end{eqnarray}
It would be convenient to let
\begin{eqnarray} \epsilon=\left(\begin{array}{c}
\epsilon^\prime\\
\epsilon^{\prime\prime} \end{array}\right).\end{eqnarray}

From(\ref{basic}), we can deduce that
\begin{equation}\label{commu}
[\Gamma^{(1)},\Gamma^{(2)}]\epsilon=0,
\end{equation}
which leads to
\begin{eqnarray}\label{realcommu1}
&&(\tilde{E}E_{1}\Gamma_{23}+E_{2}\Gamma_{03}-\tilde{E}E_{2}\Gamma_{13}
-\tilde{E}B_{1}\Gamma_{01}+B_{2}\Gamma_{12}-\tilde{E}B_{2}\Gamma_{02}
\nonumber\\
&&+B_{3}\Gamma_{13}-\tilde{E}B_{3}\Gamma_{03}-\Gamma_{23}+(E_{1}B_{1}+E_{2}B_{2})\Gamma_{01})\epsilon'=0,
\end{eqnarray}
and
\begin{eqnarray}\label{realcommu2}
&&(-\tilde{E}E_{1}\Gamma_{23}+E_{2}\Gamma_{03}+\tilde{E}E_{2}\Gamma_{13}
+\tilde{E}B_{1}\Gamma_{01}+B_{2}\Gamma_{12}+\tilde{E}B_{2}\Gamma_{02}
\nonumber\\
&&+B_{3}\Gamma_{13}+\tilde{E}B_{3}\Gamma_{03}+\Gamma_{23}-(E_{1}B_{1}+E_{2}B_{2})\Gamma_{01})\epsilon''=0.
\end{eqnarray}
In order to have nonzero solution to the equation (\ref{commu}), one
of  the equations (\ref{realcommu1}), (\ref{realcommu2}) must  have
nonzero solution.

Let
\begin{equation}
A\epsilon'=0
\end{equation}
denote the equation (\ref{realcommu1}). If the equation
\begin{equation}\label{square}
A^{2}\epsilon'=0
\end{equation}
do not have nonzero solutions, the equation (\ref{realcommu1})
also do not have nonzero solutions. The equation (\ref{square})
gives
\begin{eqnarray}\label{realsquare}
0&=&[-(\tilde{E}E_{1}-1)^{2}+(\tilde{E}B_{1}-E_{1}B_{1}-E_{2}B_{2})^{2}
+(E_{2}-\tilde{E}B_{3})^{2}-B_{2}^{2}
\nonumber\\
&&\quad -(\tilde{E}E_{2}-B_{3})^{2}+\tilde{E}^{2}B_{2}^{2}]\epsilon'
\nonumber\\
&&+2[(\tilde{E}E_{1}-1)(-\tilde{E}B_{1}+E_{1}B_{1}+E_{2}B_{2})
+B_{2}(E_{2}-\tilde{E}B_{3})
\nonumber\\
&&\quad -\tilde{E}B_{2}(\tilde{E}E_{2}-B_{3})]\Gamma_{0123}\epsilon'
.
\end{eqnarray}

Because $(\Gamma_{0123})^{2}=-\textrm{I}$, so $\Gamma_{0123}$ only
have eigenvalues $\pm i$.  Thus the necessary condition for
equation (\ref{realsquare}) to have nonzero solutions is the
constant term and the coefficient of $\Gamma_{0123}$ on its right
hand side  must be zero simultaneously. This gives us two
equations
\begin{eqnarray}\label{condition}
0&=&-(\tilde{E}E_{1}-1)^{2}+(1-\tilde{E}^{2})(E_{2}^{2}-B_{2}^{2}-B_{3}^{2})+(\tilde{E}B_{1}-E_{1}B_{1}-E_{2}B_{2})^{2},
\nonumber\\
0&=&(\tilde{E}E_{1}-1)(\tilde{E}B_{1}-E_{1}B_{1}-E_{2}B_{2})-(1-\tilde{E}^{2})E_{2}B_{2},
\end{eqnarray}
which should hold simultaneously. The similar analysis on the
equation (\ref{realcommu2}) leads to the same conditions
(\ref{condition}). If the conditions (\ref{condition}) cannot be
satisfied by the fluxes, the equations
(\ref{realcommu1}),(\ref{realcommu2}) have no nonzero solution, so
the configurations can not be supersymmetric. In other words, the
equations in (\ref{condition}) are the necessary condition for
supersymmetry. Moreover, the fluxes must respect the inequities
(\ref{defcon1}), (\ref{defcon2}).

In Appendix A, we prove that (\ref{condition}) only have two
solutions which do not break (\ref{defcon1}) and (\ref{defcon2}).
\begin{itemize} \item One solution is
\begin{eqnarray}\label{solution1}
-1&<&\tilde{E}=E_{1}<1,\hspace{3ex}B_{1}\ne
\frac{E_{1}E_{2}B_{2}}{1-E_{1}^{2}},\hspace{3ex}
1<E_{1}^{2}+E_{2}^{2}<1+B_{3}^{2},
\nonumber\\
B_{2}&=&\pm\sqrt{\frac{(1-E_{1}^{2})(1-E_{1}^{2}-E_{2}^{2}+B_{3}^{2})}{E_{1}^{2}+E_{2}^{2}-1}}.
\end{eqnarray}
\item The other solution is
\begin{eqnarray}\label{solution2}
-1&<&\tilde{E}=E_{1}<1,\hspace{3ex}B_{1}\ne
\frac{E_{1}B_{2}}{E_{2}},\hspace{3ex}B_{3}=0,
\nonumber\\
E_{2}&=&\pm\sqrt{1-E_{1}^{2}}.
\end{eqnarray}
\end{itemize}

These two solutions are just the necessary conditions for
supersymmetric configurations. We will show that they are also
sufficient conditions if we choose right sign for $B_{1}-
\frac{E_{1}E_{2}B_{2}}{1-E_{1}^{2}}$ or $B_{1}-
\frac{E_{1}B_{2}}{E_{2}}$ . As the first step to prove the
sufficiency, we will directly use $\Gamma$ matrix method to study
two simple cases. We will show that the general solutions
(\ref{solution1},\ref{solution2}) could be related to these two
simple cases via T-duality and Lorentz transformation. In these
two simple cases, we will let $\tilde{E}=E_{1}=0$, and let
$B_{2}=0$ in the second case (\ref{solution2}). Now the equation
(\ref{commu}) is
\begin{eqnarray}\label{coso1}
\left(\begin{array}{cc}
m_{11}&0\\
0&m_{22}
\end{array}\right)\epsilon
=0.
\end{eqnarray}
where
\begin{eqnarray}
m_{11}\equiv
E_{2}\Gamma_{03}+B_{2}\Gamma_{12}+B_{3}\Gamma_{13}-\Gamma_{23}+E_{2}B_{2}\Gamma_{01},
\nonumber\\
m_{22}\equiv
E_{2}\Gamma_{03}+B_{2}\Gamma_{12}+B_{3}\Gamma_{13}+\Gamma_{23}-E_{2}B_{2}\Gamma_{01}.
\end{eqnarray}

\begin{itemize}
\item[i)]Case 1: $\tilde{E}=E_{1}=0$, $B_3\neq 0$, other fluxes
satisfy (\ref{solution1})\\
In this case,  one can multiply
\begin{displaymath}
\left(\begin{array}{cc}
\Gamma_{13}&0\\
0&\Gamma_{13}
\end{array}\right)
\end{displaymath}
on the equation (\ref{coso1}) and obtain
\begin{equation}
M\epsilon=\epsilon,
\end{equation}
where $M$ is
\begin{displaymath}
\frac{1}{B_{3}}\left(\begin{array}{cc}
E_{2}\Gamma_{01}+B_{2}\Gamma_{23}+\Gamma_{12}-E_{2}B_{2}\Gamma_{03}&0\\
0&E_{2}\Gamma_{01}+B_{2}\Gamma_{23}-\Gamma_{12}+E_{2}B_{2}\Gamma_{03}
\end{array}\right)
\end{displaymath}

Since the equations
\begin{eqnarray}\label{MGamma}
\Gamma^{(1)}\epsilon=\epsilon, \hspace{5ex}
 M\epsilon=\epsilon
\end{eqnarray}
imply that
\begin{equation}
\Gamma^{(2)}\epsilon=\left\{\begin{array}{ll}
\epsilon,&\hspace{5ex} \mbox{if
$B_1>0$}\\
-\epsilon,&\hspace{5ex} \mbox{if $B_{1}<0$}\end{array}\right.
\end{equation}
 the condition (\ref{basic}) now is equivalent to (\ref{MGamma})
 with $B_1>0$ for D1-D3 or $B_1<0$ for D1-$\bar{\textrm{D}3}$
 system. It is easy to check that
\begin{eqnarray}
M^{2}=\textrm{I},\ \textrm{Tr}M=0,
\end{eqnarray}
\begin{eqnarray}
[M,\Gamma^{(1)}]=0,\   \textrm{Tr}(M\Gamma^{(1)})=0,
\end{eqnarray}
and
\begin{eqnarray}
[M,\tilde{\Gamma}_{11}]=0, \ \textrm{Tr}(M\tilde{\Gamma}_{11})=0.
\end{eqnarray}
From these properties, we can conclude that when the fluxes
satisfy (\ref{solution1}) with $\tilde{E}=E_{1}=0,B_{1}>0$, D1-D3
system preserve $1/4$ supersymmetries, and when the fluxes satisfy
(\ref{solution1}) and $\tilde{E}=E_{1}=0,B_{1}<0$,
D1-$\bar{\textrm{D}3}$ system preserve $1/4$ supersymmetries.

\item[ii)] Case 2: $\tilde{E}=E_{1}=0$, $B_2= 0$, other fluxes
satisfy (\ref{solution2})\\
 In
this case, $B_2=B_3=0$,  so the relation (\ref{commu}) is simplified
to
\begin{eqnarray}\label{coso2}
\left(\begin{array}{cc}
E_{2}\Gamma_{03}-\Gamma_{23}&0\\
0&E_{2}\Gamma_{03}+\Gamma_{23}
\end{array}\right)\epsilon=0.
\end{eqnarray}
Let
\begin{displaymath}
\left(\begin{array}{cc}
\Gamma_{23}&0\\
0&-\Gamma_{23}
\end{array}\right)
\end{displaymath}
multiply the equation (\ref{coso2}), we obtain
\begin{eqnarray}
N\epsilon=\epsilon.
\end{eqnarray}
where
\begin{eqnarray}
N=-E_{2}\left(\begin{array}{cc}
\Gamma_{02}&0\\
0&-\Gamma_{02}\nonumber
\end{array}\right)
\end{eqnarray}
Similarly, we also find that $\Gamma^{(1)}\epsilon=\epsilon$ and
$N\epsilon=\epsilon$ imply
\begin{equation}
\Gamma^{(2)}\epsilon=\left\{\begin{array}{ll}
\epsilon,&\hspace{5ex} \mbox{if
$B_1>0$}\\
-\epsilon,&\hspace{5ex} \mbox{if $B_{1}<0$}\end{array}\right.
\end{equation}
Similar to the matrix $M$ above, $N$ satisfy
\begin{eqnarray}
N^{2}=\textrm{I},\ \textrm{Tr}N=0,
\end{eqnarray}
\begin{eqnarray}
[N,\Gamma^{(1)}]=0, \  \textrm{Tr}(N\Gamma^{(1)})=0,
\end{eqnarray}
and
\begin{eqnarray}
[N,\tilde{\Gamma}_{11}]=0, \ \textrm{Tr}(N\tilde{\Gamma}_{11})=0.
\end{eqnarray}
Therefore, when the fluxes satisfy (\ref{solution2}) and
$\tilde{E}=E_{1}=0,\ B_{2}=0,\ B_{1}>0$, D1-D3 system preserve
$1/4$ supersymmetries, and when the fluxes satisfy
(\ref{solution2}) and $\tilde{E}=E_{1}=0,\ B_{2}=0,\ B_{1}<0$,
D1-$\bar{\textrm{D}3}$ system preserve $1/4$ supersymmetries.
\end{itemize}

With the above detailed analysis of two simple supersymmetric
configurations, let us turn to the general solutions
(\ref{solution1}) and (\ref{solution2}). The key point is that
since $E_1=\tilde{E}_1$ both solutions could be related to the
above two simple cases via T-duality and Lorentz transformation.
We leave the details of the transformation to Appendix B and just
give the final result here. The D1-D3(or $\bar{\textrm{D}3}$)
system with the fluxes satisfying (\ref{solution1}) is actually
equivalent to D1-D3(or $\bar{\textrm{D}3}$) system without
electric field on D1 worldvolume and
 \begin{eqnarray}
F_{D3}=\frac{1}{2\pi\alpha'}\left(\begin{array}{cccc}
0&0&-\hat{E}_{2}&0\\
0&0&\hat{B}_{3}&-\hat{B}_{2}\\
\hat{E}_{2}&-\hat{B}_{3}&0&\hat{B}_{1}\\
0&\hat{B}_{2}&-\hat{B}_{1}&0
\end{array}\right)
\end{eqnarray}
on D3 worldvolume, with $\hat{E}_2,\hat{B}_1,\hat{B}_2$ and
$\hat{B}_3$ being given in (\ref{hat1}). This is the same
configuration we have discussed before. Besides the solution
(\ref{solution1}), the extra requirement for supersymmetry is
$B_{1}-\frac{E_{1}E_{2}B_{2}}{1-E_{1}^{2}}>0$ for D1-D3 or
$B_{1}-\frac{E_{1}E_{2}B_{2}}{1-E_{1}^{2}}<0$ for
D1-$\bar{\textrm{D}3}$. Furthermore, since the electric field
along $X^1$ direction vanishes, one can do one more T-duality
along $X^1$ to get a D0-D2 system with fluxes. And another
T-duality leads to the equivalent supersymmetric intersecting
D1-D1 configurations with relative angle and motion.

For the solutions (\ref{solution2}), the similar treatment shows
that the solutions are also sufficient condition for supersymmetry
provided that $B_1-\frac{E_1B_2}{E_2}>0$ for D1-D3 or
$B_1-\frac{E_1B_2}{E_2}<0$ for D1-$\bar{\textrm{D}3}$. Similarly
the configurations could be related to fluxed D0-D2 system and
intersecting D1-D1 at angle with relative motion. All the details
on T-duality and equivalence with other configurations could be
found in Appendix B.

In summary, we have proved that \begin{itemize} \item when the
fluxes satisfy (\ref{solution1}) and
$B_{1}-\frac{E_{1}E_{2}B_{2}}{1-E_{1}^{2}}>0$, or when the fluxes
satisfy (\ref{solution2}) and $B_{1}-\frac{E_{1}B_{2}}{E_{2}}>0$,
D1-D3 systems are supersymmetric. \item when the fluxes satisfy
(\ref{solution1}) and
$B_{1}-\frac{E_{1}E_{2}B_{2}}{1-E_{1}^{2}}<0$, or when the fluxes
satisfy (\ref{solution2}) and $B_{1}-\frac{E_{1}B_{2}}{E_{2}}<0$,
D1-$\bar{\textrm{D3}}$ systems are supersymmetric. \item The
supersymmetric D1-D3 configurations we have found keep one-quarter
supersymmetries and are dual to the supersymmetric D1-D1 systems
studied in \cite{chen}.
\end{itemize}

\section{Open String Quantization and Pair Creation}

In this section, we will study generic nonsupersymmetric
configurations. We will discuss the open string excitations between
D1 and D3(or $\bar{\textrm{D3}}$)-branes by doing quantization of
the open string with boundary conditions, which are determined by
the fluxes on the D-branes. The excitation modes could be real but
not integer or half-integer, and even could also be complex. There
are various interesting issues to address. We will mainly focus on
the open string pair production and mass spectrum of near-BPS
configurations.

As usual, the boundary conditions at the ends of the open string
decide the modes expansion. In our case, the boundary conditions at
two ends are different. At $\sigma=0$ endpoint, we have boundary
condition:
\begin{eqnarray}\label{13bcl}
0&=&\partial_{\sigma}X^{0}+\tilde{E}\partial_{\tau}X^{1},
\nonumber\\
0&=&\partial_{\sigma}X^{1}+\tilde{E}\partial_{\tau}X^{0},
\nonumber\\
0&=&\partial_{\tau}X^{2},
\nonumber\\
0&=&\partial_{\tau}X^{3}.
\end{eqnarray}
While at $\sigma=\pi$ endpoint, we have
\begin{eqnarray}\label{13bcr}
0&=&\partial_{\sigma}X^{0}+E_{1}\partial_{\tau}X^{1}+E_{2}\partial_{\tau}X^{2},
\nonumber\\
0&=&\partial_{\sigma}X^{1}+E_{1}\partial_{\tau}X^{0}
+B_{3}\partial_{\tau}X^{2}-B_{2}\partial_{\tau}X^{3},
\nonumber\\
0&=&\partial_{\sigma}X^{2}+E_{2}\partial_{\tau}X^{0}
-B_{3}\partial_{\tau}X^{1}+B_{1}\partial_{\tau}X^{3},
\nonumber\\
0&=&\partial_{\sigma}X^{3}+B_{2}\partial_{\tau}X^{1}-B_{1}\partial_{\tau}X^{2}.
\end{eqnarray}

Without losing generality, we let
$X^{2}\vert_{\sigma=0}=X^{3}\vert_{\sigma=0}=0$. In $4,5,\cdots,9$
directions, the usual Dirichlet boundary conditions are imposed.
We let the distance of two branes  be $y$ in $x^4$ direction.

Now we do mode expansions for $X^{0},X^{1},X^{2},X^{3}$ with ansatz
\begin{eqnarray}\label{ucmode}
X^{\mu}&=&x_{0}^{\mu}+B^{\mu}_{0}\sigma-C_{0}^{\mu}\tau
\nonumber\\
&&
+\sum_{r=n+A}\frac{ia^{\mu}_{r}}{r}(\textrm{e}^{-ir(\tau-\sigma)}+\textrm{e}^{-ir(\tau+\sigma)})
\nonumber\\
&&+\sum_{r=n+A}\frac{ib^{\mu}_{r}}{r}(\textrm{e}^{-ir(\tau-\sigma)}-\textrm{e}^{-ir(\tau+\sigma)})
\nonumber\\
&&+\cdots,
\end{eqnarray}
where "$\cdots$" denote all possible other modes.

Imposing the boundary conditions (\ref{13bcl}),(\ref{13bcr}) to
(\ref{ucmode}), we have
\begin{eqnarray}\label{premodeeq}
b^{0}_{r}&=&\tilde{E}a^{1}_{r},
\nonumber\\
b^{1}_{r}&=&\tilde{E}a^{0}_{r},
\nonumber\\
a^{2}_{r}&=&a^{3}_{r}=0
\end{eqnarray}
and
\begin{eqnarray}\label{modeeq}
0&=&(1-\tilde{E}E_{1})(1-\textrm{e}^{-i2\pi A})a^{0}_{r}
+(\tilde{E}-E_{1})(1+\textrm{e}^{-i2\pi A})a^{1}_{r}
\nonumber\\
&&-E_{2}(1-\textrm{e}^{-i2\pi A})b_{r}^{2},
\nonumber\\
0&=&(\tilde{E}-E_{1})(1+\textrm{e}^{-i2\pi A})a^{0}_{r}
+(1-\tilde{E}E_{1})(1-\textrm{e}^{-i2\pi A})a^{1}_{r}
\nonumber\\
&&-B_{3}(1-\textrm{e}^{-i2\pi
A})b_{r}^{2}+B_{2}(1-\textrm{e}^{-i2\pi A})b_{r}^{3},
\nonumber\\
0&=&[-E_{2}(1+\textrm{e}^{-i2\pi A})
+\tilde{E}B_{3}(1-\textrm{e}^{-i2\pi A})]a^{0}_{r}
+[B_{3}(1+\textrm{e}^{-i2\pi A})
\nonumber\\
& &-\tilde{E}E_{2}(1-\textrm{e}^{-i2\pi A})]a^{1}_{r}
+(1+\textrm{e}^{-i2\pi A})b_{r}^{2} -B_{1}(1-\textrm{e}^{-i2\pi
A})b_{r}^{3},
\nonumber\\
0&=&-\tilde{E}B_{2}(1-\textrm{e}^{-i2\pi A})a^{0}_{r}
-B_{2}(1+\textrm{e}^{-i2\pi A})a^{1}_{r} +B_{1}(1-\textrm{e}^{-i2\pi
A})b_{r}^{2}
\nonumber\\
&&+(1+\textrm{e}^{-i2\pi A})b_{r}^{3}.
\end{eqnarray}
If some fields really have $r$ modes , the coefficient matrix of
(\ref{modeeq}) must have zero determinant. This help us to fix $A$
from the equation
\begin{eqnarray}
&&(B_{1}-\tilde{E}E_{1}B_{1}-\tilde{E}E_{2}B_{2})^{2}\textrm{tan}^{4}\pi
A
\nonumber\\
&& +[-1+E^{2}_{2}-B_{2}^{2}-B_{3}^{2}+(E_{1}B_{1}+E_{2}B_{2})^{2}
\nonumber\\
&&\quad +\tilde{E}^{2}(-E_{1}^{2}-E_{2}^{2}
+B_{1}^{2}+B_{2}^{2}+B_{3}^{2})+2\tilde{E}E_{1}
\nonumber\\
&&\quad
-2\tilde{E}B_{1}(E_{1}B_{1}+E_{2}B_{2})]\textrm{tan}^{2}\pi A
\nonumber\\
&&-(\tilde{E}-E_{1})^{2}=0
\end{eqnarray}
It is easy to check that when the fluxes satisfy (\ref{solution1})
or (\ref{solution2}), there exist only integer modes.

For simplicity, we let $\tilde{E}=E_{1}$ and other fluxes be free.
This include the supersymmetric case, and also include many other
nonsupersymmetric ones. Now the possible values for $A$ are
\begin{eqnarray} A=0
\end{eqnarray}
which gives the integer modes, and also
\begin{eqnarray}\label{rorcmode}
(\textrm{tan} \pi A)^{2}=\frac{\Lambda}{\Delta^2},
\end{eqnarray}
where
 \begin{eqnarray}
  \Lambda&=&(1-E^{2}_{1}-E^{2}_{2})(1-E^{2}_{1}+B^{2}_{2})
+(1-E^{2}_{1})B^{2}_{3}\label{D}\\
 \Delta&=& (1-E^{2}_{1})B_{1}-E_{1}E_{2}B_{2}.\label{delta}
 \end{eqnarray}
  If $\Lambda>0$, there will be real fractional excitation modes. We will discuss this case in
subsection 3.2. If $\Lambda<0$,
then $A$ is pure imaginary. 
Let us discuss this case first. In this case, it is convenient to
introduce a real parameter $\epsilon$,
\begin{equation}\label{epsilondef}
\epsilon\equiv\frac{1}{\pi}\textrm{arctanh}\frac{\sqrt{-\Lambda}}{\Delta}.
\end{equation}
The sign of $\epsilon$ is the same as the sign of $\Delta \neq 0$.

The mode expansions of $X^\mu$ and its super-partner is quite
involved. From them one can define the symplectic form to do
quantization. After proper linear transformation, one can write the
Hamiltonian in a canonical way. The details on the mode expansion
and quantization could be found in Appendix C.

The Hamiltonian of 1-3 string in $0,1,2,3$ directions is
\setlength\arraycolsep{2pt}
\begin{eqnarray}
H_{(0,1,2,3)}=&\frac{1}{2}\int_{0}^{\pi}\textrm{d}\sigma&
(\partial_{\tau}X^{\mu}\partial_{\tau}X_{\mu}+\partial_{\sigma}X^{\mu}\partial_{\sigma}X_{\mu}
\nonumber\\
&&\
+i\psi_{+}^{\mu}\partial_{\sigma}\psi_{\mu+}-i\psi_{-}^{\mu}\partial_{\sigma}\psi_{\mu-}).
\end{eqnarray}
Here $\mu=0,1,2,3$.

Due to the existence of non-diagonal terms, the Hamiltonian looks
messy in terms of the original independent modes. In terms of the
transformed modes,  we obtain
\begin{eqnarray}\label{newham}
H_{0,1,2,3}&=&H_{0-mode}+\frac{1}{2}\sum_{n\ne0}c_{n}c_{-n}+\frac{1}{2}\sum_{n\ne0}d_{n}d_{-n}
\nonumber\\
&&
-\frac{1}{2}\sum_{n}b_{n+i\epsilon}b_{-n-i\epsilon}
-\frac{1}{2}\sum_{n}b_{n-i\epsilon}b_{-n+i\epsilon}
\nonumber\\
&&-\frac{1}{2}\sum_{r}r\phi_{r}\phi_{-r}-\frac{1}{2}\sum_{r}r\xi_{r}\xi_{-r}
\nonumber\\
&&+\frac{1}{2}\sum_{r}(r+i\epsilon)\beta_{r+i\epsilon}\beta_{-r-i\epsilon}
+\frac{1}{2}\sum_{r}(r-i\epsilon)\beta_{r-i\epsilon}\beta_{-r+i\epsilon}.
\end{eqnarray}
In (\ref{newham}), $H_{0-mode}$ comes from the zero-mode
\begin{eqnarray}
H_{0-mode}&\equiv&-\frac{\pi}{2}(1-E_{1}^{2}-E_{2}^{2})(C_{0}^{0})^{2}
+\frac{\pi}{2}(1-E_{1}^{2}+B_{2}^{2}+B_{3}^{2})(C_{0}^{1})^{2}
\nonumber\\
&&-\pi E_{2}B_{3}C_{0}^{0}C_{0}^{1},
\end{eqnarray}
and the (anti-)commutation relations between modes take the
canonical form:
\begin{eqnarray}
[c_{n},c_{m}]&=&n\delta_{n,-m},\hspace{3ex}
[d_{n},d_{m}]=n\delta_{n,-m},\hspace{3ex} [c_{n},d_{m}]=0,
\nonumber\\
\lbrack b_{n+i\epsilon},
b_{m-i\epsilon}\rbrack&=&-(n+i\epsilon)\delta_{n,-m}.
\nonumber\\
\lbrace\phi_{r},\phi_{s}\rbrace&=&\delta_{r,-s},\hspace{3ex}
\lbrace\xi_{r},\xi_{s}\rbrace=\delta_{r,-s},\hspace{3ex}
\lbrace\phi_{r},\xi_{s}\rbrace=0,
\nonumber\\
\lbrace\beta_{r+i\epsilon},\beta_{s-i\epsilon}\rbrace&=&-\delta_{r,-s}.
\end{eqnarray}

 Written in normal order, the Hamiltonian is
\begin{eqnarray}\label{normal}
H_{0,1,2,3}&=&H_{0-mode}+\sum_{n>0}c_{-n}c_{n}+\sum_{n>0}d_{-n}d_{n}
-\sum_{n\geqslant0}b_{-n-i|\epsilon|}b_{n+i|\epsilon|}
\nonumber\\
&&-\sum_{n>0}b_{-n+i|\epsilon|}b_{n-i|\epsilon|}
+\sum_{r>0}r\phi_{-r}\phi_{r}+\sum_{r>0}r\xi_{-r}\xi_{r}
\nonumber\\
&&-\sum_{r\geqslant0}(r+i|\epsilon|)\beta_{-r-i|\epsilon|}\beta_{r+i|\epsilon|}
-\sum_{r>0}(r-i|\epsilon|)\beta_{-r+i|\epsilon|}\beta_{r-i|\epsilon|}
\nonumber\\
&&+\mathcal {E}_{0}.
\end{eqnarray}
where $\mathcal {E}_{0}$ is the zero-point energy
\begin{eqnarray}
\mathcal {E}_{0}=\left\{\begin{array}{cc}
0,&\quad \textrm{R sector}\ ,\\
\frac{i|\epsilon|}{2}-\frac{1}{4},&\quad \textrm{NS sector}\ .
\end{array}\right.
\end{eqnarray}

 Taking into account of the excitations along
other directions and ghosts, the vacuum state in NS sector
$|0\rangle_{NS}$ has energy
\begin{eqnarray}
\mathcal
{E}_{v}=\frac{y^{2}}{2\pi}+\frac{i|\epsilon|}{2}-\frac{1}{2}=\frac{y^{2}}{2\pi}+\left\{\begin{array}{cc}
\frac{i\epsilon}{2}-\frac{1}{2}&\quad \Delta>0\ ,\\
-\frac{i\epsilon}{2}-\frac{1}{2}&\quad \Delta<0\ .
\end{array}\right.
\end{eqnarray}
where $\Delta$ was defined as (\ref{delta}).

The GSO projection is quite subtle in the cases with background
fluxes. There is spectral flow when the fluxes are
varied\cite{ChenB, Witten99}. In our case, when $\Delta<0$, the
GSO projection on $|0\rangle_{NS}$ is different from the case when
$\Delta>0$. From (\ref{bomore}), (\ref{femore}), we find that
$\beta^{\nu\pm}_{r\pm i|\epsilon|}\ (\nu=0,1,3)$ have different
relations with $\beta_{r\pm i|\epsilon|}$ when the signs of
$\Delta$ are different. Thus when $\Delta$ change its sign, the
normal ordering in (\ref{normal}) indicates  that the orientation
of D-brane has changed. As discussed in \cite{ChenB},  the
eigenvalues of GSO projection operator on $|0\rangle_{NS}$ could
be defined by  the function $\frac{1+f(\Delta)}{2}$, where
$f(\Delta)$ can only be $\pm 1$, and must take opposite values
when $\Delta$ changes sign. In D1-D3 systems, we will prove that
\begin{eqnarray}\label{nobarf}
f(\Delta)=\left\{\begin{array}{cc}
-1&\quad \Delta>0\ ,\\
1&\quad \Delta<0
\end{array}\right.
\end{eqnarray}
 in subsection 3.2. In D1-$\bar{\textrm{D3}}$ systems,
$f(\Delta)$ take opposite values to (\ref{nobarf}).

\subsection{Open String Pair Creation}

Now we can calculate 1-loop vacuum amplitude $\mathcal {A}$ for
open strings between D1 and D3 (or $\bar{\textrm{D3}}$) branes:
\begin{eqnarray}\label{oneloop}
\mathcal
{A}&=&i\frac{\varphi_{flux}}{\alpha'^{2}}V_{2}\int_{0}^{\infty}dt
\frac{q^{\frac{y^{2}}{4\pi^{2}\alpha'}}}{t^{2}\eta^{9}(it)\theta_{1}(|\epsilon|t|it)}\times
\nonumber\\
&&\times[\theta_{3}^{3}(0|it)\theta_{3}(|\epsilon|t|it)
+f(\Delta)\lambda\theta_{4}^{3}(0|it)\theta_{4}(|\epsilon|t|it)
\nonumber\\
&&\quad-\theta_{2}^{3}(0|it)\theta_{2}(|\epsilon|t|it)]\ ,
\end{eqnarray}
where we restore the dependence on $2\pi\alpha'$. In the above
relation,  $q=\textrm{e}^{-2\pi t}$,
$\theta_{i}(\nu|\tau),i=1,2,3,4$ are theta functions, and
$\eta(\tau)$ is Dedekind eta function. The parameter $\lambda$
characterize the GSO projection, being $1$ for D3 or $-1$ for
$\bar{\textrm{D}3}$. The $\varphi_{flux}$ is a real algebraic
function of fluxes,
\begin{eqnarray}\label{varphi}
\varphi_{flux}\equiv
\frac{D}{64\pi^{4}[E_{1}(1-E_{1}^{2}-E_{2}^{2}+B_{2}^{2}+B_{3}^{2})-E_{2}B_{1}B_{2}]}\
,
\end{eqnarray}
where $D$ is defined in (\ref{ddef}). The $\varphi_{flux}$ comes
from several sources. One part of it is from the integral of
zero-modes, because $C_{0}^{\mu}=-2\alpha' p^{\mu},\ \mu=0,1$. And
we need to multiply factor $\frac{i}{[x^{0},x^{1}]}$ on $V_{2}$ in
$\mathcal {A}$ for a noncommutative normalization. This
normalization factor contributes to $\varphi_{flux}$. The other
numerical factors to $\varphi_{flux}$ come from orientation,
integral measure and GSO projection.

When there exist complex modes, there are open string pairs
production. The creation rate $\omega$ could be read from the 1-loop
vacuum amplitude\cite{Bachas}
\begin{eqnarray}\label{omegadef}
\omega&=&-2\textrm{Im}(\frac{i}{V_{2}}\mathcal{A})
=-\frac{2\mathcal {A}}{V_{2}}
\end{eqnarray}
Since there exist poles in the integrand of $\mathcal{A}$ at
$t=\frac{l}{|\epsilon|} (l=1,2,\cdots)$, the
 contour integration give us nonvanishing $\omega$. When $\Delta>0$,
\begin{eqnarray}\label{rateresult}
\omega&=&-\frac{\varphi_{flux}}{\alpha'^{2}}\sum_{l=1}^{\infty}\frac{|\epsilon|}{l^{2}}
\textrm{e}^{-\frac{y^{2}l}{2\pi\alpha'
|\epsilon|}}\frac{1}{\eta^{12}(i\frac{l}{|\epsilon|})} \nonumber\\
&&\times\left\{\begin{array}{cc}
((-1)^{l}-1)\theta_{2}^{4}(0|i\frac{l}{|\epsilon|})&\textrm{D}1-\textrm{D}3\ ,\\
((-1)^{l}-1)\theta_{3}^{4}(0|i\frac{l}{|\epsilon|})
+((-1)^{l}+1)\theta_{4}^{4}(0|i\frac{l}{|\epsilon|})&\textrm{D}1-\bar{\textrm{D}3}\
.
\end{array}\right.
\end{eqnarray}
When $\Delta<0$, the conclusion is opposite.

When $|\epsilon|$ are very small, from the asymptotic behavior of
$\theta$ and $\eta$ functions, we have
\begin{eqnarray}
\frac{((-1)^{l}-1)\theta_{2}^{4}(0|i\frac{l}{|\epsilon|})}{\eta^{12}(i\frac{l}{|\epsilon|})}
\thicksim 16((-1)^{l}-1)(1+O(\textrm{e}^{-\frac{2\pi
l}{|\epsilon|}}))
\end{eqnarray}
and
\begin{eqnarray}
&&\frac{((-1)^{l}-1)\theta_{3}^{4}(0|i\frac{l}{|\epsilon|})
+((-1)^{l}+1)\theta_{4}^{4}(0|i\frac{l}{|\epsilon|})}{\eta^{12}(i\frac{l}{|\epsilon|})}
\nonumber\\
&&\thicksim 2\textrm{e}^{\frac{\pi
l}{|\epsilon|}}(-1)^{l}(1+O(\textrm{e}^{-\frac{\pi
l}{|\epsilon|}})).
\end{eqnarray}
Therefore, we learn that when two branes are far away from each
other, the  contribution from $l=1$ dominate. When two branes move
to each other, the contribution from higher values $l$ become more
and more important. Since $|\epsilon|\approx 0$, for D1-D3 case
with $\Delta >0$, the open string pair production is exponentially
suppressed. This is consistent with the fact that the system is
now near-BPS. On the other hand, for D1-$\bar{\textrm{D}3}$ with
$D>0$, if $y$ is finite, it may suppress the creation of open
string pair production, while if $y \approx 0$, the pair creation
is enhanced, especially for large $l$. This indicates that the
system is far from being supersymmetric.

\subsection{GSO Projection and Near Massless States}

In this subsection, we will determine the GSO projection in NS
sector, and prove (\ref{nobarf}). We will also study the spectrum
of open strings in the near-BPS case. This happens when the
excitation modes are real and the fluxes are taken to be in a
decoupling limit.

In last subsection, we find that the eigenvalue of GSO projection
operator on $|0\rangle_{NS}$ is a function $\frac{1+f(\Delta)}{2}$,
in which $f(\Delta)$  can only be $\pm 1$. The value of $f(\Delta)$
 depends on the sign of $\Delta$, and take opposite values when
$\Delta$ change sign. To determine GSO projection, we analyze
$\epsilon\to 0$ limit of one-loop amplitude $\mathcal {A}$.  From
the definition of $\epsilon$, we know that $D\to 0$, so
$\varphi_{flux}\to 0$ at the zero $\epsilon$ limit. However, at the
same time $\theta_{1}(|\epsilon|t|it) \to 0$ too. Using the
definition of theta function, we get (for D1-D3)
\begin{eqnarray}
\lim_{\epsilon\to 0}\mathcal
{A}&=&i\frac{|\Delta|V_{2}}{128\pi^{4}\alpha'^{2}[E_{1}(1-E_{1}^{2}-E_{1}^{2}+B_{2}^{2}+B_{3}^{2})-E_{2}B_{1}B_{2}]}
\times
\nonumber\\
&&\times\int_{0}^{\infty}dt
\frac{q^{\frac{y^{2}}{4\pi^{2}\alpha'}}}{t^{3}\eta^{12}(it)}[\theta_{3}^{4}(0|it)
+f(\Delta)\theta_{4}^{4}(0|it) -\theta_{2}^{4}(0|it)]\  .
\end{eqnarray}
In section 2 and Appendix B, we have already obtained all
supersymmetric conditions for D1-D3 systems with fluxes. These
supersymmetric conditions are equivalent to
\begin{eqnarray}
D&=&0\ ,
\nonumber\\
\Delta&>&0\ .
\end{eqnarray}
In the supersymmetric case, one-loop amplitude must be zero. So
$\lim_{\epsilon\rightarrow 0}\mathcal {A}$ must be zero when
$\Delta>0, D=0$. Recall that theta function satisfies Jacobi's
`abstruse identity'
\begin{eqnarray} \theta_{3}^{4}(0|it) -\theta_{4}^{4}(0|it)
-\theta_{2}^{4}(0|it)=0\ ,
\end{eqnarray}
so $f(\Delta)$ must be $-1$ when $\Delta>0$. This is just
(\ref{nobarf}). Similarly, we can determine the GSO projection in
D1-$\bar{\textrm{D3}}$, which is just changing $f(\Delta)$ to
$-f(\Delta)$.

Let us discuss the case when $\tilde{E}=E_{1}$ and $D>0$. From
(\ref{rorcmode}), we know that there are real fractional modes in
this case. Let us introduce  real parameters
\begin{eqnarray}\label{dtildedef}
\tilde{D}\equiv\sqrt{\Lambda}\ ,
\end{eqnarray}
and
\begin{eqnarray}\label{tildeepsdef}
\tilde{\epsilon}\equiv\frac{1}{\pi}\textrm{arctan}\frac{\tilde{D}}{\Delta}\
.
\end{eqnarray}
We can obtain the mode expansions and relations of modes like
(\ref{bosmoex}), (\ref{fermoex}), (\ref{bomore}) and
(\ref{femore}), by replacing $i\epsilon,\ D$ with
$\tilde{\epsilon},\ -i\tilde{D}$ respectively.

The one-loop amplitude $\tilde{\mathcal {A}}$ is now
\begin{eqnarray}\label{tildeoneloop}
\tilde{\mathcal
{A}}&=&\frac{\tilde{\varphi}_{flux}}{\alpha'^{2}}V_{2}\int_{0}^{\infty}dt
\frac{q^{\frac{y^{2}}{4\pi^{2}\alpha'}}}{t^{2}\eta^{9}(it)\theta_{1}(-i|\tilde{\epsilon}|t|it)}\times
\nonumber\\
&&\times[\theta_{3}^{3}(0|it)\theta_{3}(-i|\tilde{\epsilon}|t|it)
+f(\Delta)\lambda\theta_{4}^{3}(0|it)\theta_{4}(-i|\tilde{\epsilon}|t|it)
\nonumber\\
&&\quad-\theta_{2}^{3}(0|it)\theta_{2}(-i|\tilde{\epsilon}|t|it)]\
,
\end{eqnarray}
where
\begin{eqnarray}\label{varphi} \tilde{\varphi}_{flux}\equiv
\frac{\tilde{D}}{64\pi^{4}[E_{1}(1-E_{1}^{2}-E_{2}^{2}+B_{2}^{2}+B_{3}^{2})-E_{2}B_{1}B_{2}]}\
.
\end{eqnarray}
$f(\Delta)$ is defined as (\ref{nobarf}) too. The function in the
integrand of (\ref{tildeoneloop}) now are pure imaginary and only
have pole at $t=0$ on positive real axis. Thus there is no open
string pair production in this case.

In this case, it is meaningful to discuss the mass spectrum of
open string between D1 and D3-brane. In D1-D3 systems, if we let
$y=0$, the ground state in NS sector has energy
\begin{eqnarray}
\mathcal {E}_{v}=\frac{|\tilde{\epsilon}|}{2}-\frac{1}{2}\ .
\end{eqnarray}
When $\Delta>0$, the excited states $\beta_{-\frac{1}{2}\pm
\tilde{\epsilon}}|0\rangle_{NS}$ have energies
$-\frac{\tilde{\epsilon}}{2}$ and $\frac{3\tilde{\epsilon}}{2}$
respectively.  Now under GSO projection, the ground state
$|0\rangle_{NS}$ is projected out while  $\beta_{-\frac{1}{2}\pm
\tilde{\epsilon}}|0\rangle_{NS}$ survive. When $\tilde{\epsilon}$
is very small, these states become near massless. Other than this,
the states
$\beta^{\mu}_{-\frac{1}{2}}|0\rangle_{NS}\equiv\beta^{\mu+}_{-\frac{1}{2}}|0\rangle_{NS}
,\ \mu=4,5,\cdots ,9 $ all have energies
$\frac{\tilde{\epsilon}}{2}$, they are all near massless when
$\tilde{\epsilon}$ is very small. Furthermore, one can act on
these states with an arbitrary polynomial consisting of
$b_{-\tilde{\epsilon}}$ with energy $\tilde{\epsilon}$. This
action gives rise to a large number of near massless states.

The configurations with $\tilde{\epsilon} \approx 0$ are called
near-BPS. This happens when $\tilde{D} \approx 0$ or
$\Delta\rightarrow \infty$. The solutions of $\tilde{D}=0$ are
(\ref{solution1}) and (\ref{solution2}), which are the
supersymmetric conditions. It is expected that when $\tilde{D}
\approx 0$ there are many near massless states. On the other hand,
the fact that the case with $\Delta\rightarrow \infty$ has many
near massless states sounds strange. One way to understand this
fact is to take a large $B_1$ to get a large $\Delta$. Effectively
we can neglect other fluxes and simplify our system to D1-D3 with
a large magnetic field $B_1$. For this simplified system, it has
been known to be near-BPS and has many near-massless
states\cite{ChenB}. In this case, the magnetic field on $D3$ may
induce a large number D1's so the system is near-BPS. In fact, one
can understand this configuration from a dual description in
matrix model\cite{Minwalla}.

However, still for  D1-D3 systems, when $\Delta<0$, the picture is
very different. Because now the ground state $|0\rangle_{NS}$
survive GSO projection and the first excited states are all
projected out, the tachyon is there though we take
$\tilde{\epsilon}\to 0$ limit. Now the system is far from
supersymmetry.

For D1-$\bar{\textrm{D3}}$ systems, the conclusions are opposite.
When $\Delta<0$, there are many near massless states when
$\tilde{\epsilon}\to 0$. And when $\Delta>0$, there is no state
become near massless when $\tilde{\epsilon}\to 0$.

\section{Conclusions and Discussions}

In this paper, we studied D1-D3 (or $\bar{\textrm{D3}}$) systems
with constant fluxes in flat spacetime. We worked out all
configurations which keep one-quarter supersymmetries. The result
were summarized at the end of section 2. The supersymmetric
configurations are T-dual to the D0-D2 brane system, and dual to
supersymmetric intersecting D1-D1 with relative angle and motion,
which has been studied in \cite{chen}. Furthermore, we
investigated generic nonsupersymmetric configurations by
quantizing the open string between D1 and D3 (or
$\bar{\textrm{D3}}$). In general, the open string modes could be
complex or real fractional, rather than integer or half-integer.
When the modes are complex, we obtained the open string pair
production rate from 1-loop amplitude. When the modes are real, we
discussed the open string mass spectrum and found that there could
exist a large number of near massless states when the system is
near-BPS. This is reminiscent of the same phenomenon discussed in
\cite{ChenB,Witten99}.

Our study of fluxed D1-D3 (or $\bar{\textrm{D3}}$) system shows
that turning on background fluxes can recover the supersymmetries
of a non-BPS system. Generically speaking, it is quite difficult
to decide if a system with fluxes is supersymmetric or not, since
the supersymmetry analysis is quite involved. In particular, when
the dimensionality of D-brane  gets large, the number of possible
background fluxes are large so that it is not easy to work out all
the supersymmetric configurations.  It would be interesting to
find a more effective way to solve the problem.

From our study, it turned out that the GSO projection is quite
subtle in the study of open string excitation, especially when
there are background fluxes. We decide the GSO projection by
taking the BPS limit. This would be a nice way to determine the
GSO projection in more general setting.

In the study of the open string spectrum, we noticed that there
would be large number of light states if the system is near-BPS.
One way to reach near-BPS configuration is to let the magnetic
field in the codimendion be large. Effectively one can neglect
other fluxes in the system and the system is reduced to the ones
studied in \cite{ChenB}. This picture will be true for other
systems.

In this article, we do not discuss the case that the fluxes take
the critical values. It was found in \cite{newn} that when the
electric field take the critical value, one can define a novel
string theory. This string theory is an interacting open string
theory, in which the close strings decouple from the open ones.
 In our case, if we let
$\tilde{E}=E_{1}=1$, we are actually discussing the fluxed D3-branes
in a noncommutative open string theory.  Naively, from
(\ref{rorcmode}), it seems that only integer bosonic modes can
exist, which indicates that the configurations is supersymmetric no
matter what kind of fluxes we turn on the D3-brane. There would be
no open string pair production or tachyon condensation, as the case
discussed in other non-BPS system with critical electric
field\cite{ChenSun}.
We look forward to a rigorous discussion about the critical fluxes
in this system.

\section*{Acknowledgments}

The work was partially supported by NSFC Grant No. 10535060,
10775002, and NKBRPC (No. 2006CB805905).

\begin{appendix}
\section{Solutions of (\ref{condition})}

 Here we will analyze the possible solutions of
(\ref{condition}). Firstly, the second equation of
(\ref{condition}) can be factorized into
\begin{equation}\label{A}
(\tilde{E}-E_{1})[(\tilde{E}E_{1}-1)B_{1}+\tilde{E}E_{2}B_{2}]=0.
\end{equation}
One solution is $\tilde{E}=E_{1}$. If so,  the first equation of
(\ref{condition}) is
\begin{equation}\label{a}
-(1-E_{1}^{2})^{2}+(1-E_{1}^{2})(E_{2}^{2}-B_{2}^{2}-B_{3}^{2})+E_{2}^{2}B_{2}^{2}=0.
\end{equation}
\begin{enumerate}
\item If $1-E_{1}^{2}-E_{2}^{2}\ne 0$, we obtain
(\ref{solution1}). ($B_{1}\ne \frac{E_{1}E_{2}B_{2}}{1-E_{1}^{2}}$
come from (\ref{defcon2}))\item  If $1-E_{1}^{2}-E_{2}^{2}=0$,
Eq.(\ref{a}) now is
\begin{equation}
(1-E_{1}^{2})B_{3}^{2}=0.
\end{equation}
Since we require $1-E_{1}^{2}\ne 0$, so $B_{3}=0$. This is
(\ref{solution2}). ($B_{1}\ne \frac{E_{1}B_{2}}{E_{2}}$ come from
(\ref{defcon2}))
\end{enumerate}

When $\tilde{E}\ne E_{1}$, from (\ref{A}),
\begin{equation}\label{b}
(\tilde{E}E_{1}-1)B_{1}+\tilde{E}E_{2}B_{2}=0.
\end{equation}

If $\tilde{E}E_{1}-1=0$, we must let $E_{2}=0$ or $B_{2}=0$. If
$\tilde{E}E_{1}-1=0,E_{2}=0$, the first equation of
(\ref{condition}) is
\begin{equation}
-(1-\frac{1}{E_{1}^{2}})(B_{2}^{2}+B_{3}^{2})+(E_{1}-\frac{1}{E_{1}})^{2}B_{1}^{2}=0,
\end{equation}
which require
\begin{equation}
B_{2}^{2}+B_{3}^{2}=(E_{1}^{2}-1)B_{1}^{2}.
\end{equation}
Then the left hand side of (\ref{defcon2}) equal to $1-E_{1}^{2}$.
From (\ref{defcon1}) and $\tilde{E}E_{1}-1=0$, we know
$1-E_{1}^{2}<0$, so (\ref{defcon2}) cannot be satisfied. To let
$B_{2}=0$ lead to the  same conclusion. Therefore there is no
solution when $\tilde{E}\ne E_{1},\tilde{E}E_{1}-1=0$.

When $\tilde{E}\ne E_{1}, \tilde{E}E_{1}-1\ne 0$, we deduce from
(\ref{b}) that
\begin{equation}\label{c}
B_{1}=\frac{\tilde{E}E_{2}B_{2}}{1-\tilde{E}E_{1}}.
\end{equation}
Because (\ref{defcon1}),(\ref{defcon2}), we obtain
\begin{eqnarray}
1-\tilde{E}^{2}&>&0,
\nonumber\\
E_{2}^{2}-B_{2}^{2}-B_{3}^{2}&<&1-E_{1}^{2}+B_{1}^{2}-(E_{1}B_{1}+E_{2}B_{2})^{2}
\end{eqnarray}
So the right hand side of the first equation of (\ref{condition})
satisfy
\begin{eqnarray}\label{d}
&&-(\tilde{E}E_{1}-1)^{2}+(1-\tilde{E}^{2})(E_{2}^{2}-B_{2}^{2}-B_{3}^{2})+(\tilde{E}B_{1}-E_{1}B_{1}-E_{2}B_{2})^{2}
\nonumber\\
&&<-(\tilde{E}E_{1}-1)^{2}+(1-\tilde{E}^{2})(1-E_{1}^{2}+B_{1}^{2}-(E_{1}B_{1}+E_{2}B_{2})^{2})
\nonumber\\
&&\quad +(\tilde{E}B_{1}-E_{1}B_{1}-E_{2}B_{2})^{2}
\nonumber\\
&&=-(\tilde{E}E_{1}-1)^{2}+(1-\tilde{E}^{2})(1-E_{1}^{2})
\nonumber\\
&&\quad
+(1-\tilde{E}^{2})(B_{1}^{2}-(E_{1}B_{1}+E_{2}B_{2})^{2})+(\tilde{E}B_{1}-(E_{1}B_{1}+E_{2}B_{2}))^{2}
\nonumber\\
&&=-(\tilde{E}-E_{1})^{2}+(B_{1}-\tilde{E}(E_{1}B_{1}+E_{2}B_{2}))^{2}.
\end{eqnarray}
From (\ref{c}), we obtain
\begin{equation}
\tilde{E}(E_{1}B_{1}+E_{2}B_{2})=B_{1}.
\end{equation}
So
\begin{eqnarray}\label{e}
&&-(\tilde{E}-E_{1})^{2}+(B_{1}-\tilde{E}(E_{1}B_{1}+E_{2}B_{2}))^{2}
\nonumber\\
&&=-(\tilde{E}-E_{1})^{2}<0.
\end{eqnarray}
From (\ref{d}) and (\ref{e}), we learn that  the first equation of
(\ref{condition}) can not be satisfied. So we prove that
(\ref{defcon1}),(\ref{defcon2}) are in contradiction with
(\ref{condition}) when $\tilde{E}\ne E_{1}, \tilde{E}E_{1}-1\ne
0$.

In summary, when $\tilde{E}\ne E_{1}$, (\ref{condition}) have no
solutions which do not break (\ref{defcon1}) and (\ref{defcon2}).

Therefore, the solutions (\ref{solution1}) and (\ref{solution2})
are all possible solutions of (\ref{condition}) which obey
(\ref{defcon1}) and (\ref{defcon2}).

\section{T-dual Discussions}

T-duality is a powerful technique for the study of D-branes. The
different D-brane systems could be related to each other by
T-duality. Shortly speaking, it exchange Neumann and Dirichlet
boundary conditions\cite{book1,book2} of open string. One nice
property of T-duality is that it keeps supersymmetry.

For a string ending on D-branes with fluxes, the boundary
conditions is \cite{seiwit}
\begin{equation}
G_{\mu\nu}\partial_{\sigma}X^{\nu}+iF_{\mu\nu}\partial_{t}X^{\nu}=0.
\end{equation}
For general $\textrm{D}1-\textrm{D}3$ system with fluxes
(\ref{d1flux}) and (\ref{d3flux}), the boundary conditions of
string ending on $\textrm{D}1$ are
\begin{eqnarray}\label{0d1bc1}
\partial_{\sigma}X^{0}+i\tilde{E}\partial_{t}X^{1}=0,
\nonumber\\
\partial_{\sigma}X^{1}+i\tilde{E}\partial_{t}X^{0}=0,
\end{eqnarray}
and the boundary conditions of string ending on $\textrm{D}3$ are
\begin{eqnarray}\label{0d3bc1}
0&=&\partial_{\sigma}X^{0}+iE_{1}\partial_{t}X^{1}+iE_{2}\partial_{t}X^{2},
\nonumber\\
0&=&\partial_{\sigma}X^{1}+iE_{1}\partial_{t}X^{0}
+\textrm{i}B_{3}\partial_{t}X^{2}-iB_{2}\partial_{t}X^{3},
\nonumber\\
0&=&\partial_{\sigma}X^{2}+iE_{2}\partial_{t}X^{0}
-iB_{3}\partial_{t}X^{1}+iB_{1}\partial_{t}X^{3},
\nonumber\\
0&=&\partial_{\sigma}X^{3}+iB_{2}\partial_{t}X^{1}-iB_{1}\partial_{t}X^{2},
\end{eqnarray}

Now let us do T-dual in $X^{1}$ direction. The boundary conditions
(\ref{0d1bc1}) and (\ref{0d3bc1}) are changed to
\begin{eqnarray}\label{0d1bc2}
0&=&\partial_{\sigma}(X^{0}-\tilde{E}X^{1}) ,
\nonumber\\
0&=&\partial_{t}(X^{1}-\tilde{E}X^{0}),
\end{eqnarray}
and
\begin{eqnarray}\label{0d3bc2}
0&=&\partial_{\sigma}(X^{0}-E_{1}X^{1})+iE_{2}\partial_{t}X^{2},
\nonumber\\
0&=&\partial_{t}(X^{1}-E_{1}X^{0} -B_{3}X^{2}+B_{2}X^{3}),
\nonumber\\
0&=&\partial_{\sigma}(X^{2}+B_{3}X^{1})+iE_{2}\partial_{t}X^{0}
+iB_{1}\partial_{t}X^{3},
\nonumber\\
0&=&\partial_{\sigma}(X^{3}-B_{2}X^{1})-iB_{1}\partial_{t}X^{2}.
\end{eqnarray}

Let us first consider the solution (\ref{solution1}). Defining
\begin{eqnarray}
X^{0'} = \frac{X^{0}-E_{1}X^{1}}{\sqrt{1-E_{1}^{2}}},\hspace{5ex}
X^{1'} = \frac{X^{1}-E_{1}X^{0}}{\sqrt{1-E_{1}^{2}}},
\end{eqnarray}
then we get the boundary conditions
\begin{eqnarray}\label{1d1bc1}
0=\partial_{\sigma}X^{0'},\hspace{5ex} 0=\partial_{t}X^{1'},
\end{eqnarray}
and
\begin{eqnarray}\label{1d3bc2}
0&=&\partial_{\sigma}X^{0'}+i\hat{E}_{2}\partial_{t}X^{2},
\nonumber\\
0&=&\partial_{t}(X^{1'}-\hat{B}_{3}X^{2}+\hat{B}_{2}X^{3}),
\nonumber\\
0&=&\partial_{\sigma}(X^{2}+\hat{B}_{3}X^{1'})+i\hat{E}_{2}\partial_{t}X^{0'}
+i\hat{B}_{1}\partial_{t}X^{3},
\nonumber\\
0&=&\partial_{\sigma}(X^{3}-\hat{B}_{2}X^{1'})-i\hat{B}_{1}\partial_{t}X^{2},
\end{eqnarray}
where
\begin{eqnarray}\label{hat1}
\hat{E}_{2}&\equiv&\frac{E_{2}}{\sqrt{1-E_{1}^{2}}}\ , \quad
\hat{B}_{1}\equiv B_{1}-\frac{E_{1}E_{2}B_{2}}{1-E_{1}^{2}}\ ,
\nonumber\\
\hat{B}_{2}&\equiv&\frac{B_{2}}{\sqrt{1-E_{1}^{2}}}\ , \quad
\hat{B}_{3}\equiv\frac{B_{3}}{\sqrt{1-E_{1}^{2}}}\ .
\end{eqnarray}
After doing T-duality on $X^{1'}$, we come back to D1-D3 system
but now
with fluxes
\begin{equation}\label{hatrela2}
\hat{\tilde{E}}=\hat{E}_{1}=0,\
 1<\hat{E}_{2}^{2}<1+\hat{B}_{3}^{3},\
\hat{B}_{1} \ne 0,\
\hat{B}_{2}^{2}=\frac{1-\hat{E}_{2}^{2}+\hat{B}_{3}^{2}}{\hat{E}_{2}^{2}-1}
\end{equation}
In section 2, we prove this system is supersymmetric if
$\hat{B_{1}}>0$. Because T-duality and Lorentz transformation do not
change the number of  supersymmetries, we conclude that with fluxes
satisfy (\ref{solution1}) and
$B_{1}-\frac{E_{1}E_{2}B_{2}}{1-E_{1}^{2}}=\hat{B_{1}}>0$, the
original D1-D3 system preserve $\frac{1}{4}$ supersymmetry.

Similar discussions can do for D1-$\bar{\textrm{D}3}$ systems. We
find that with fluxes constrained by (\ref{solution1}) and
$B_{1}-\frac{E_{1}E_{2}B_{2}}{1-E_{1}^{2}}<0$,
D1-$\bar{\textrm{D3}}$ system preserve $\frac{1}{4}$
supersymmetry.

Furthermore for D1-D3, if we  do rotation
\begin{eqnarray}\label{rot}
X^{1''}&=&\frac{1}{\sqrt{1+\hat{B}_{2}^{2}+\hat{B}_{3}^{2}}}(X^{1'}-\hat{B}_{3}X^{2}+\hat{B}_{2}X^{3}),
\nonumber\\
X^{2'}&=&\frac{1}{\sqrt{1+\hat{B}_{3}^{2}}}(\hat{B}_{3}X^{1'}+X^{2}),
\nonumber\\
X^{3'}&=&\frac{\sqrt{\hat{E}_{2}^{2}-1}}{\hat{E}_{2}\hat{B}_{3}\sqrt{1+\hat{B}_{3}^{2}}}
(-\hat{B}_{2}X^{1'}+\hat{B}_{2}\hat{B}_{3}X^{2}+(1+\hat{B}_{3}^{2})X^{3})
\end{eqnarray}
(\ref{1d3bc2}) become
\begin{eqnarray}\label{1d3bc3}
0&=&\partial_{\sigma}X^{0'}+i\frac{\hat{E}_{2}}{\sqrt{1+\hat{B}_{3}^{2}}}\partial_{t}X^{2'}
-i\frac{\hat{B}_{2}\sqrt{\hat{E}_{2}^{2}-1}}{\sqrt{1+\hat{B}_{3}^{2}}}\partial_{t}X^{3'},
\nonumber\\
0&=&\partial_{t}X^{1''},
\nonumber\\
0&=&\partial_{\sigma}X^{2'}+i\frac{\hat{E}_{2}}{\sqrt{1+\hat{B}_{3}^{2}}}\partial_{t}X^{0'}
+i\frac{\hat{B}_{1}\sqrt{\hat{E}_{2}^{2}-1}}{\hat{E}_{2}\hat{B}_{3}}\partial_{t}X^{3'},
\nonumber\\
0&=&\partial_{\sigma}X^{3'}-i\frac{\hat{B}_{2}\sqrt{\hat{E}_{2}^{2}-1}}{\sqrt{1+\hat{B}_{3}^{2}}}\partial_{t}X^{0'}
-i\frac{\hat{B}_{1}\sqrt{\hat{E}_{2}^{2}-1}}{\hat{E}_{2}\hat{B}_{3}}\partial_{t}X^{2'}.
\end{eqnarray}
The system now becomes static D0-D2 system with constant fluxes.
One can do another T-duality in $X^{2'}$ direction and change the
system to intersecting D1-D1's with relative angle and motion.

In \cite{chen}, the D2-D2 system with generic fluxes has been
studied. The supersymmetric configurations found there could be
dual to two intersecting D1's, which are moving relative to each
other with angle. The supersymmetric condition is
\begin{equation}\label{chencondition}
-e^{2}_{2}(1-\beta^{2}_{1})(1-\beta^{2}_{2})+\sin^{2}\theta=\beta^{2}_{1}+\beta^{2}_{2}-2\beta_{1}\beta_{2}\cos
\theta.
\end{equation}
In this equation, $e_{2}$ is the electric flux on the second D1,
$\beta_{1},\beta_{2}$ are normal speed of two D1s, $\theta$ is the
angle between two strings.

In our case, after T-duality in $X^{2'}$ direction, the system now
is a D1-D1 system with \cite{myers}
\begin{eqnarray}\label{1d1d1}
\beta_{1}&=&0,\
\beta_{2}=\frac{\sin\theta\hat{E}_{2}}{\sqrt{1+\hat{B}_{3}^{2}}},\
e_{1}=0,\
e_{2}=-\frac{\sin\theta\hat{B}_{2}\sqrt{\hat{E}_{2}^{2}-1}}{\sqrt{1-\beta_{2}^{2}}\sqrt{1+\hat{B}_{3}^{2}}},
\nonumber\\
\cot\theta&=&-\frac{\hat{B}_{1}\sqrt{\hat{E}_{2}^{2}-1}}{\hat{E}_{2}\hat{B}_{3}},
\end{eqnarray}
where $e_{1},e_{2},\beta_{1},\beta_{2},\theta$ have the same
meaning as the ones in (\ref{chencondition}). It is easy to check
that the above identifications (\ref{1d1d1}) satisfy the
supersymmetric condition (\ref{chencondition}). This confirms that
our supersymmetric analysis is correct.

For the solutions (\ref{solution2}), let
\begin{eqnarray}
X^{0'}=\frac{X^{0}-E_{1}X^{1}}{\sqrt{1-E_{1}^{2}}}, \hspace{3ex}
X^{1'}=\frac{X^{1}-E_{1}X^{0}}{\sqrt{1-E_{1}^{2}}},
\end{eqnarray}
the boundary conditions (\ref{0d1bc2}) and (\ref{0d3bc2}) can be
rewritten as
\begin{eqnarray}\label{2d1bc1}
0=\partial_{\sigma}X^{0'},\hspace{5ex} 0=\partial_{t}X^{1'},
\end{eqnarray}
and
\begin{eqnarray}\label{2d3bc1}
0&=&\partial_{\sigma}X^{0'}\pm i\partial_{t}X^{2},
\nonumber\\
0&=&\partial_{t}(X^{1'}\pm\frac{B_{2}}{E_{2}}X^{3}),
\nonumber\\
0&=&\partial_{\sigma}X^{2}\pm
i\partial_{t}X^{0'}+i(B_{1}-\frac{E_{1}B_{2}}{E_{2}})\partial_{t}X^{3},
\nonumber\\
0&=&\partial_{\sigma}(X^{3}\mp\frac{B_{2}}{E_{2}}X^{1'})-i(B_{1}-\frac{E_{1}B_{2}}{E_{2}})\partial_{t}X^{2}.
\end{eqnarray}
Similarly, we find that this system is T-dual to D1-D3 system with
fluxes
\begin{eqnarray}
\hat{\tilde{E}}&=&\hat{E}_{1}=0,\
\hat{B}_{1}=B_{1}-\frac{E_{1}B_{2}}{E_{2}},\ \hat{B}_{2}=0,\
\hat{B}_{3}=0,
\nonumber\\
 \hat{E}_{2}&=&\pm\sqrt{1-\hat{E}_{1}^{2}}.
\end{eqnarray}
We know this system preserve $\frac{1}{4}$ supersymmetry from
section 2. Thus if the fluxes satisfy (\ref{solution2}) and
$B_{1}-\frac{E_{1}B_{2}}{E_{2}}>0$, the original D1-D3 system
preserve $\frac{1}{4}$ supersymmetry. And with fluxes as
(\ref{solution2}) and $B_{1}-\frac{E_{1}B_{2}}{E_{2}}<0$,
D1-$\bar{\textrm{D3}}$ system preserve $\frac{1}{4}$
supersymmetry.

For D1-D3 systems, if we do rotation
\begin{eqnarray}
X^{1''}&=&\frac{X^{1'}\pm\frac{B_{2}}{E_{2}}X^{3}}{\sqrt{1+\frac{B_{2}^{2}}{E_{2}^{2}}}},
\nonumber\\
X^{3'}&=&\frac{X^{3}\mp\frac{B_{2}}{E_{2}}X^{1'}}{\sqrt{1+\frac{B_{2}^{2}}{E_{2}^{2}}}}.
\end{eqnarray}
(\ref{2d3bc1}) equal to
\begin{eqnarray}\label{2d3bc2}
0&=&\partial_{\sigma}X^{0'}\pm i\partial_{t}X^{2},
\nonumber\\
0&=&\partial_{t}X^{1''},
\nonumber\\
0&=&\partial_{\sigma}X^{2}\pm
i\partial_{t}X^{0'}+i\frac{1}{\sqrt{1+\frac{B_{2}^{2}}{E_{2}^{2}}}}(B_{1}
-\frac{E_{1}B_{2}}{E_{2}})\partial_{t}X^{3'},
\nonumber\\
0&=&\partial_{\sigma}X^{3'}-i\frac{1}{\sqrt{1+\frac{B_{2}^{2}}{E_{2}^{2}}}}(B_{1}
-\frac{E_{1}B_{2}}{E_{2}})\partial_{t}X^{2}.
\end{eqnarray}
This system becomes a D0-D2 system with fluxes.

We now do another T-duality in $X^{3'}$ direction for the D0-D2
system mentioned above. The system becomes intersecting D1-D1
system with
\begin{eqnarray}\label{2d1d1}
\beta_{1}=\beta_{2}=0,\  e_{1}=0,\ e_{2}=\sin\theta,\
\cot\theta=\frac{1}{\sqrt{1+\frac{B_{2}^{2}}{E_{2}^{2}}}}(B_{1}
+\frac{E_{1}B_{2}}{E_{2}}).
\end{eqnarray}
It is easy to see that (\ref{2d1d1}) satisfy the supersymmetric
condition (\ref{chencondition}).

\section{Mode expansion and quantization}

When $\Lambda<0$, the mode expansion for $X^{1},X^{2},X^{3}$ with
all possible modes are
\begin{eqnarray}\label{bosmoex}
X^{\mu}&=&x_{0}^{\mu}+B^{\mu}_{0}\sigma-C_{0}^{\mu}\tau
+\sum_{n\ne 0}
\frac{ia^{\mu}_{n}}{n}(\textrm{e}^{-in(\tau-\sigma)}+\textrm{e}^{-in(\tau+\sigma)})
\nonumber\\
&&+\sum_{n\ne 0}
\frac{ib^{\mu}_{n}}{n}(\textrm{e}^{-in(\tau-\sigma)}-\textrm{e}^{-in(\tau+\sigma)})
\nonumber\\
&&+\sum_{n+i\epsilon}\frac{ia^{\mu}_{n+i\epsilon}}{n+i\epsilon}
(\textrm{e}^{-i(n+i\epsilon)(\tau-\sigma)}+\textrm{e}^{-i(n+i\epsilon)(\tau+\sigma)})
\nonumber\\
&&+\sum_{n+i\epsilon} \frac{ib^{\mu}_{n+i\epsilon}}{n+i\epsilon}
(\textrm{e}^{-i(n+i\epsilon)(\tau-\sigma)}-\textrm{e}^{-i(n+i\epsilon)(\tau+\sigma)})
\nonumber\\
&&+\sum_{n-i\epsilon}\frac{ia^{\mu}_{n-i\epsilon}}{n-i\epsilon}
(\textrm{e}^{-i(n-i\epsilon)(\tau-\sigma)}+\textrm{e}^{-i(n-i\epsilon)(\tau+\sigma)})
\nonumber\\
&&+\sum_{n-i\epsilon} \frac{ib^{\mu}_{n-i\epsilon}}{n-i\epsilon}
(\textrm{e}^{-i(n-i\epsilon)(\tau-\sigma)}-\textrm{e}^{-i(n-i\epsilon)(\tau+\sigma)})
\end{eqnarray}

Not all the coefficients of these modes are nonzero or
independent. From the boundary conditions (\ref{premodeeq}) and
(\ref{modeeq}), we can find that there are following relations
\begin{eqnarray}\label{bomore}
x_{0}^{2}&=&x_{0}^{3}=C_{0}^{2}=C_{0}^{3}=a_{n}^{2}=a_{n}^{3}=a_{n\pm
i\epsilon}^{2}=a_{n\pm i\epsilon}^{3}=0.
\nonumber\\
B_{0}^{0}&=&E_{1}C_{0}^{1},\ B_{0}^{1}=E_{1}C_{0}^{0},\
B_{0}^{2}=E_{2}C_{0}^{0}-B_{3}C_{0}^{1},\
B_{0}^{3}=B_{2}C_{0}^{1}.
\nonumber\\
b_{n}^{0}&=&E_{1}a_{n}^{1},\ b_{n}^{1}=E_{1}a_{n}^{0},\
b_{n}^{2}=E_{2}a_{n}^{0}-B_{3}a_{n}^{1},\
b_{n}^{3}=B_{2}a_{n}^{1}.
\nonumber\\
b_{n\pm i\epsilon}^{0}&=&E_{1}a_{n\pm i\epsilon}^{1},\ b_{n\pm
i\epsilon}^{1}=E_{1}a_{n\pm i\epsilon}^{0}
\nonumber\\
a_{n\pm i\epsilon}^{0}&=&\frac{E_{2}}{1-E_{1}^{2}}b_{n\pm
i\epsilon}^{2},\ a_{n\pm
i\epsilon}^{1}=\frac{(1-E_{1}^{2})B_{3}\mp
B_{2}D}{(1-E_{1}^{2})(1-E_{1}^{2}+B_{2}^{2})}b_{n\pm
i\epsilon}^{2},
\nonumber\\
b_{n\pm i\epsilon}^{3}&=&\frac{B_{2}B_{3}\pm
D}{1-E_{1}^{2}+B_{2}^{2}}b_{n\pm i\epsilon}^{2}.
\end{eqnarray}
Here $D$ is defined as
\begin{eqnarray}\label{ddef}
D\equiv\sqrt{-\Lambda}\ .
\end{eqnarray}
And
$x^{0}_{0},x^{1}_{0},C_{0}^{0},C_{0}^{1},a_{n}^{0},a_{n}^{1},b_{n\pm
i\epsilon}^{2}$ are independent. We know $D$ is real since
$\Lambda<0$.

The mode expansion for the fermions can be obtained similary. The
possible mode expansions are
\begin{eqnarray}\label{fermoex}
\psi^{\mu}_{+}&=&\sum_{r}\alpha_{r}^{\mu+}\textrm{e}^{-ir(\tau+\sigma)}
\nonumber\\
&&+\sum_{r+i\epsilon}\beta_{r+i\epsilon}^{\mu+}\textrm{e}^{-i(r+i\epsilon)(\tau+\sigma)}
+\sum_{r-i\epsilon}\beta_{r-i\epsilon}^{\mu+}\textrm{e}^{-i(r-i\epsilon)(\tau+\sigma)},
\nonumber\\
\psi^{\mu}_{-}&=&\sum_{r}\alpha_{r}^{\mu-}\textrm{e}^{-ir(\tau-\sigma)}
\nonumber\\
&&+\sum_{r+i\epsilon}\beta_{r+i\epsilon}^{\mu-}\textrm{e}^{-i(r+i\epsilon)(\tau-\sigma)}
+\sum_{r-i\epsilon}\beta_{r-i\epsilon}^{\mu-}\textrm{e}^{-i(r-i\epsilon)(\tau-\sigma)}.
\end{eqnarray}
The coefficients of these modes have the following relations
\begin{eqnarray}\label{femore}
\alpha_{r}^{0-}&=&\frac{(1+E_{1}^{2})\alpha_{r}^{0+}+2E_{1}\alpha_{r}^{1+}}{1-E_{1}^{2}},
\alpha_{r}^{1-}=\frac{(1+E_{1}^{2})\alpha_{r}^{1+}+2E_{1}\alpha_{r}^{0+}}{1-E_{1}^{2}},
\nonumber\\
\alpha_{r}^{2+}&=&-\alpha_{r}^{2-}=-\frac{(E_{2}-E_{1}B_{3})\alpha_{r}^{0+}+(E_{1}E_{2}-B_{3})\alpha_{r}^{1+}}{1-E_{1}^{2}}
\nonumber\\
\alpha_{r}^{3+}&=&-\alpha_{r}^{3-}=-\frac{E_{1}B_{2}\alpha_{r}^{0+}+B_{2}\alpha_{r}^{1+}}{1-E_{1}^{2}},
\nonumber\\
\beta_{r\pm i\epsilon}^{0+}&=&(\frac{E_{1}(1-E_{1}^{2})B_{3}\mp
E_{1}B_{2}D}{(1-E_{1}^{2})(1-E_{1}^{2}+B_{2}^{2})}
-\frac{E_{2}}{1-E_{1}^{2}})\beta_{r\pm i\epsilon}^{2+},
\nonumber\\
\beta_{r\pm i\epsilon}^{0-}&=&(-\frac{E_{1}(1-E_{1}^{2})B_{3}\mp
E_{1}B_{2}D}{(1-E_{1}^{2})(1-E_{1}^{2}+B_{2}^{2})}
-\frac{E_{2}}{1-E_{1}^{2}})\beta_{r\pm i\epsilon}^{2+},
\nonumber\\
\beta_{r\pm
i\epsilon}^{1+}&=&(\frac{E_{1}E_{2}}{1-E_{1}^{2}}-\frac{(1-E_{1}^{2})B_{3}\mp
B_{2}D}{(1-E_{1}^{2})(1-E_{1}^{2}+B_{2}^{2})} )\beta_{r\pm
i\epsilon}^{2+},
\nonumber\\
\beta_{r\pm
i\epsilon}^{1-}&=&(-\frac{E_{1}E_{2}}{1-E_{1}^{2}}-\frac{(1-E_{1}^{2})B_{3}\mp
B_{2}D}{(1-E_{1}^{2})(1-E_{1}^{2}+B_{2}^{2})} )\beta_{r\pm
i\epsilon}^{2+},
\nonumber\\
\beta_{r\pm i\epsilon}^{2-}&=&-\beta_{r\pm i\epsilon}^{2+},
\nonumber\\
\beta_{r\pm i\epsilon}^{3+}&=&-\beta_{r\pm i\epsilon}^{3-}
=\frac{B_{2}B_{3}\pm D}{1-E_{1}^{2}+B_{2}^{2}}\beta_{r\pm
i\epsilon}^{2+}.
\end{eqnarray}
Here $r$ is integer (for R sector) or half integer (for NS
sector). Definition of $D$ can be found in (\ref{ddef}).
$\alpha_{r}^{0+},\alpha_{r}^{1+},\beta_{r\pm i\epsilon}^{2+}$ are
independent.

The symplectic form is defined as
\begin{equation}\label{omedef}
\Omega=\int_{0}^{\pi}d\sigma(\delta\Pi_{X_{\mu}}\wedge\delta
X^{\mu}-\delta\Pi_{\psi_{\mu}}\wedge\delta \psi^{\mu}),
\end{equation}
where $\psi^{\mu}=(\psi^{\mu+}, \psi^{\mu-})$ is world-sheet
Majorana spinor,  and $\Pi_{X_{\mu}},\Pi_{\psi_{\mu}}$ are the
conjugate momenta of $X^{\mu},\psi^{\mu}$
\begin{eqnarray}
\Pi_{X_{\mu}}=\eta_{\mu\nu}\partial_{\tau}X^{\mu}+(A_{\mu}^{(0)}\delta(\sigma)-A_{\mu}^{(\pi)}\delta(\sigma-\pi)),
\hspace{3ex}\Pi_{\psi_{\mu}}=\frac{i}{2}\bar{\psi}^{\nu}\gamma^{0}\eta_{\mu\nu}.
\end{eqnarray}
Here $\gamma^{0}=i\sigma^{2}$ is a two-dimensional gamma matrix.
$A_{\mu}^{(0)},A_{\mu}^{(\pi)}$ are gauge potentials on D1,D3
brane, whose field strengths are (\ref{d1flux}), (\ref{d3flux})
respectively.

With the mode expansions (\ref{bosmoex}) and (\ref{fermoex}), we
can calculate $\Omega$ using (\ref{omedef}). Because relations
(\ref{bomore}) and (\ref{femore}), we should write $\Omega$ in
independent variables at final result. After integration and some
algebraic calculations, we obtain
\begin{eqnarray}
\Omega&=&\pi(E_{1}^{2}+E_{2}^{2}-1)\delta x_{0}^{0}\wedge\delta
C_{0}^{0} +\pi(1-E_{1}^{2}+B_{2}^{2}+B_{3}^{2})\delta
x_{0}^{1}\wedge\delta C_{0}^{1}
\nonumber\\
&&-\pi E_{2}B_{3}(\delta x_{0}^{0}\wedge\delta C_{0}^{1}+\delta
x_{0}^{1}\wedge\delta C_{0}^{0})
\nonumber\\
&&+\pi^{2}(E_{1}(1-E_{1}^{2}-E_{2}^{2}+B_{2}^{2}+B_{3}^{2})-E_{2}B_{1}B_{2})\delta
C_{0}^{0}\wedge\delta C_{0}^{1}
\nonumber\\
&&+\sum_{n\ne 0}\frac{2\pi i}{n}(1-E_{1}^{2}-E_{2}^{2})\delta
a_{n}^{0}\wedge\delta a_{-n}^{0}
\nonumber\\
&&-\sum_{n\ne0}\frac{2\pi i}{n}(1-E_{1}^{2}+B_{2}^{2}
+B_{3}^{2})\delta a_{n}^{1}\wedge\delta a_{-n}^{1}
\nonumber\\
&&+\sum_{n\ne0}\frac{4\pi i}{n}E_{2}B_{3}\delta
a_{n}^{0}\wedge\delta a_{-n}^{1}
\nonumber\\
&&-\sum_{n}\frac{8\pi i}{n+i\epsilon}
(1-\frac{E_{2}^{2}}{1-E_{1}^{2}}+\frac{B_{3}^{2}}{1-E_{1}^{2}+B_{2}^{2}})
\delta b_{n+i\epsilon}^{2}\wedge \delta b_{-n-i\epsilon}^{2}
\nonumber\\
&&+\sum_{r}\pi
i(-1+\frac{(E_{2}-E_{1}B_{3})^{2}+E_{1}^{2}B_{2}^{2}}{(1-E_{1}^{2})^{2}})
\delta\alpha_{r}^{0+}\wedge\delta\alpha_{-r}^{0+}
\nonumber\\
&&+\sum_{r}\pi
i(1+\frac{(E_{1}E_{2}-B_{3})^{2}+B_{2}^{2}}{(1-E_{1}^{2})^{2}})
\delta\alpha_{r}^{1+}\wedge\delta\alpha_{-r}^{1+}
\nonumber\\
&&+\sum_{r}2\pi
i\frac{(E_{2}-E_{1}B_{3})(E_{1}E_{2}-B_{3})+E_{1}B_{2}^{2}}{(1-E_{1}^{2})^{2}}
\delta\alpha_{r}^{0+}\wedge\delta\alpha_{-r}^{1+}
\nonumber\\
&&+\sum_{r}{4\pi i}
(1-\frac{E_{2}^{2}}{1-E_{1}^{2}}+\frac{B_{3}^{2}}{1-E_{1}^{2}+B_{2}^{2}})
\delta \beta_{r+i\epsilon}^{2+}\wedge \delta
\beta_{-r-i\epsilon}^{2+}
\end{eqnarray}

From symplectic form $\Omega$, we can obtain Poisson bracket (here
is Dirac bracket) through usual way. Then we can work out result
of quantization from Dirac bracket directly. They are
\begin{eqnarray}
[x_{0}^{0},x_{0}^{1}]&=&i\frac{E_{1}(1-E_{1}^{2}-E_{2}^{2}+B_{2}^{2}+B_{3}^{2})-E_{2}B_{1}B_{2}}{D^{2}},
\nonumber\\
\lbrack
x_{0}^{0},C_{0}^{0}\rbrack&=&-i\frac{1-E_{1}^{2}+B_{2}^{2}+B_{3}^{2}}{\pi
D^{2}},
\nonumber\\
\lbrack
x_{0}^{1},C_{0}^{1}\rbrack&=&-i\frac{E_{1}^{2}+E_{2}^{2}-1}{\pi
D^{2}},
\nonumber\\
\lbrack
x_{0}^{0},C_{0}^{1}\rbrack&=&[x_{0}^{1},C_{0}^{0}]=-i\frac{E_{2}B_{3}}{\pi
D^{2}},
\nonumber\\
\lbrack C_{0}^{0},C_{0}^{1}\rbrack&=&0,
\nonumber\\
\lbrack
a_{n}^{0},a_{m}^{0}\rbrack&=&\frac{n}{4\pi}\frac{1-E_{1}^{2}+B_{2}^{2}
+B_{3}^{2}}{D^{2}}\delta_{n,-m},
\nonumber\\
\lbrack
a_{n}^{1},a_{m}^{1}\rbrack&=&\frac{n}{4\pi}\frac{E_{1}^{2}+E_{2}^{2}-1}{D^{2}}\delta_{n,-m},
\nonumber\\
\lbrack
a_{n}^{0},a_{m}^{1}\rbrack&=&\frac{n}{4\pi}\frac{E_{2}B_{3}}{D^{2}}\delta_{n,-m},
\nonumber\\
\lbrack b_{n+i\epsilon}^{2},b_{m-i\epsilon}^{2}\rbrack
&=&-\frac{n+i\epsilon}{8\pi}
\frac{(1-E_{1}^{2})(1-E_{1}^{2}+B_{2}^{2})}{D^{2}}\delta_{n,-m},
\nonumber\\
\lbrace \alpha_{r}^{0+},\alpha_{s}^{0+}\rbrace
&=&\frac{1}{2\pi}\frac{(1-E_{1}^{2})^{2}+(E_{1}E_{2}-B_{3})^{2}+B_{2}^{2}}{D^{2}}\delta_{r,-s},
\nonumber\\
\lbrace \alpha_{r}^{1+},\alpha_{s}^{1+}\rbrace
&=&\frac{1}{2\pi}\frac{-(1-E_{1}^{2})^{2}+(E_{2}-E_{1}B_{3})^{2}+E_{1}^{2}B_{2}^{2}}{D^{2}}\delta_{r,-s},
\nonumber\\
\lbrace \alpha_{r}^{0+},\alpha_{s}^{1+}\rbrace
&=&-\frac{1}{2\pi}\frac{(E_{2}-E_{1}B_{3})(E_{1}E_{2}-B_{3})+E_{1}B_{2}^{2}}{D^{2}}\delta_{r,-s},
\nonumber\\
\lbrace \beta_{r+i\epsilon}^{2+},\beta_{s-i\epsilon}^{2+}\rbrace
&=&-\frac{1}{4\pi}\frac{(1-E_{1}^{2})(1-E_{1}^{2}+B_{2}^{2})}{D^{2}}\delta_{r,-s}.
\end{eqnarray}
As expectations, $x_{0}^{0}$ and $x_{0}^{1}$ are noncommutative. A
little trouble is that $[a_{n}^{0},a_{-n}^{1}]$,\
$\{\alpha_{r}^{0+},\alpha_{-r}^{1+}\}$ are nonzero. `Non-diagonal'
(anti-)commutator will obstruct us to define Fock space. To resolve
this problem, We transfer some modes. Linear transforms we will do
for these modes can make `Non-diagonal' commutator (anti-commutator)
vanish. More over, as we have seen in section 3, those linear
transforms can transfer world-sheet Hamiltonian to an elegant form.
 Details of our linear transforms
are
\begin{eqnarray}\label{lintra}
c_{n}&\equiv&-\frac{\sqrt{\frac{4\pi
D^{2}}{1-E_{1}^{2}+B_{2}^{2}+B_{3}^{2}}}a_{n}^{0}+\sqrt{\frac{4\pi
D^{2}}{E_{1}^{2}+E_{2}^{2}-1}}a_{n}^{1}}
{\sqrt{2(1+\frac{E_{2}B_{3}}{\sqrt{(E_{1}^{2}+E_{2}^{2}-1)(1-E_{1}^{2}+B_{2}^{2}+B_{3}^{2})}})}},
\nonumber\\
d_{n}&\equiv&-\frac{\sqrt{\frac{4\pi
D^{2}}{1-E_{1}^{2}+B_{2}^{2}+B_{3}^{2}}}a_{n}^{0}-\sqrt{\frac{4\pi
D^{2}}{E_{1}^{2}+E_{2}^{2}-1}}a_{n}^{1}}
{\sqrt{2(1-\frac{E_{2}B_{3}}{\sqrt{(E_{1}^{2}+E_{2}^{2}-1)(1-E_{1}^{2}+B_{2}^{2}+B_{3}^{2})}})}},
\nonumber\\
b_{n\pm i\epsilon}&\equiv&\sqrt{\frac{8\pi D^{2}}
{(1-E_{1}^{2})(1-E_{1}^{2}+B_{2}^{2})}}b_{n\pm i\epsilon}^{2},
\nonumber\\
\phi_{r}&\equiv&\frac{\sqrt{\frac{2\pi
D^{2}}{(1-E_{1}^{2})^{2}+(E_{1}E_{2}-B_{3})^{2}+B_{2}^{2}}}\alpha_{r}^{0+}+\sqrt{\frac{2\pi
D^{2}}{-(1-E_{1}^{2})^{2}+(E_{2}-E_{1}B_{3})^{2}+E_{1}^{2}B_{2}^{2}}}\alpha_{r}^{1+}}
{\sqrt{2(1-\frac{(E_{2}-E_{1}B_{3})(E_{1}E_{2}-B_{3})+E_{1}B_{2}^{2}}
{\sqrt{((1-E_{1}^{2})^{2}+(E_{1}E_{2}-B_{3})^{2}+B_{2}^{2})
(-(1-E_{1}^{2})^{2}+(E_{2}-E_{1}B_{3})^{2}+E_{1}^{2}B_{2}^{2})}})}},
\nonumber\\
\xi_{r}&\equiv&\frac{\sqrt{\frac{2\pi
D^{2}}{(1-E_{1}^{2})^{2}+(E_{1}E_{2}-B_{3})^{2}+B_{2}^{2}}}\alpha_{r}^{0+}-\sqrt{\frac{2\pi
D^{2}}{-(1-E_{1}^{2})^{2}+(E_{2}-E_{1}B_{3})^{2}+E_{1}^{2}B_{2}^{2}}}\alpha_{r}^{1+}}
{\sqrt{2(1+\frac{(E_{2}-E_{1}B_{3})(E_{1}E_{2}-B_{3})+E_{1}B_{2}^{2}}
{\sqrt{((1-E_{1}^{2})^{2}+(E_{1}E_{2}-B_{3})^{2}+B_{2}^{2})
(-(1-E_{1}^{2})^{2}+(E_{2}-E_{1}B_{3})^{2}+E_{1}^{2}B_{2}^{2})}})}},
\nonumber\\
\beta_{r\pm i\epsilon}&\equiv&\sqrt{\frac{4\pi D^{2}}
{(1-E_{1}^{2})(1-E_{1}^{2}+B_{2}^{2})}}\beta_{r\pm
i\epsilon}^{2+}.
\end{eqnarray}

After transforms (\ref{lintra}), new operator satisfy
commute(anti-commute) relations
\begin{eqnarray}
[c_{n},c_{m}]&=&n\delta_{n,-m},\ [d_{n},d_{m}]=n\delta_{n,-m},\
[c_{n},d_{m}]=0,
\nonumber\\
\lbrack b_{n+i\epsilon},
b_{m-i\epsilon}\rbrack&=&-(n+i\epsilon)\delta_{n,-m}.
\nonumber\\
\lbrace\phi_{r},\phi_{s}\rbrace&=&\delta_{r,-s},\
\lbrace\xi_{r},\xi_{s}\rbrace=\delta_{r,-s},\
\lbrace\phi_{r},\xi_{s}\rbrace=0,
\nonumber\\
\lbrace\beta_{r+i\epsilon},\beta_{s-i\epsilon}\rbrace&=&-\delta_{r,-s}.
\end{eqnarray}

$C_{0}^{\mu}$ should transfer like $a_{n}^{\mu}$. But in this
paper, the issue we concern undergo little influence from
transforms of $C_{0}^{\mu}$. So we do not write out their
transforms explicitly.

\end{appendix}


\begin{thebibliography}{99}
\bibitem{book1} J. Polchinski, ``String Theory'', Cambridge
University Press, (1998).
\bibitem{book2} C. V. Johnson, ``D-Branes'', Cambridge
University Press, (2003).
\bibitem{supertube} D. Mateos and P. K. Townsend, ``Supertubes'', Phys.Rev.Lett. 87:011602, 2001, [arXiv:hep-th/0103030].
\bibitem{tac} D. Mateos, Selena Ng and P. K. Townsend, ``Tachyons, Supertubes and Brane/Anti-Brane
Systems'', JHEP 0203:016, 2002, [arXiv:hep-th/0112054].
\bibitem{chen} B. Chen, C. Chen and F. Lin,``Supergravity Null Scissors and Super-Crosses'',
JHEP 0304:022, 2003, [arXiv:hep-th/0303156].
\bibitem{bran} D. Bak and A. Karch, ``Supersymmetric Brane-Antibrane
Configurations'', Nucl.Phys. B626:165-182, 2002,
[arXiv:hep-th/0110039].\\
D.~s.~Bak and N.~Ohta, ``Supersymmetric D2 anti-D2 strings,''
  Phys.\ Lett.\  B {\bf 527} (2002) 131
  [arXiv:hep-th/0112034],\\
Y.~Hyakutake and N.~Ohta,
  ``Supertubes and supercurves from M-ribbons,''
  Phys.\ Lett.\  B {\bf 539} (2002) 153
  [arXiv:hep-th/0204161], \\
D.~s.~Bak, N.~Ohta and M.~M.~Sheikh-Jabbari,
  ``Supersymmetric brane anti-brane systems: Matrix model description,
  stability and decoupling limits,''
  JHEP {\bf 0209} (2002) 048
  [arXiv:hep-th/0205265],

\bibitem{myers} R. C. Myers and D. J. Winters, ``From $\textrm{D}-\bar{\textrm{D}}$ Pairs to Branes in Motion'',
JHEP 0212:061, 2002, [arXiv:hep-th/0211042].
\bibitem{zoo} D. Bak, N. Ohta and P. K. Townsend, ``The D2 Susy Zoo'',
[arXiv:hep-th/0612101].
\bibitem{Witten96}E. Witten, ``Bound states of strings and
p-branes", Nucl.Phys. B460:335-350, 1996, [hep-th/9510135].
\bibitem{Minwalla}M. Aganagic, R. Gopakumar, S. Minwalla and A.
Strominger, ``Unstable Solitons in Noncommutative Gauge Theory",
JHEP 0104:001, 2001,  [hep-th/0009142].
\bibitem{ChenB}B. Chen, H. Itoyama, T. Matsuo and K. Murakami, `` p - p-prime system with B field, branes at angles and noncommutative
geometry", Nucl.Phys.B576:177-195,2000,
[arXiv:hep-th/9910263].``World sheet and space-time properties of p
- p-prime system with b field and noncommutative geometry",
Nucl.Phys.B593:505-544,2001, [arXiv:hep-th/0005283].``Correspondence
between noncommutative soliton and open string / D-brane system via
Gaussian damping factor",
Prog.Theor.Phys.105:853-868,2001,[arXiv:hep-th/0010066].
\bibitem{Witten99}E. Witten, ``BPS Bound states of D0 - D6 and D0 - D8 systems in a B
field", JHEP 0204:012,2002,[arXiv:hep-th/0012054].
\bibitem{superd} E. Bergshoeff and P. K. Townsend, ``Super
D-branes'',  Nucl.Phys. B490:145-162, 1997, [arXiv:hep-th/9611173].
\bibitem{symmetry}  E. Bergshoeff, R. Kallosh, T. Ortin and G. Papadopoulos,
``$\kappa$-Symmetry, Supersymmetry and Intersecting Branes'',
Nucl.Phys. B502:149-169, 1997,  [arXiv:hep-th/9705040].
\bibitem{seiwit} N. Seiberg, E. Witten, ``String Theory and Noncommutative Geometry
'', JHEP 9909:032, 1999, [arXiv:hep-th/9908142].
\bibitem{Bachas}C. Bachas and M. Porrati, "Pair creation of open
string in an electric field", Phys. Lett. 296(1992)77, [arXiv:
hep-th/9209032].\\
C. Bachas, ``D-brane dynamics", Phys. Lett. B 374(2996)37,
[arXiv:hep-th/9511043]
\bibitem{pair}  J. Cho, P. Oh, C. Park and J. Shin, ``String Pair Creations in D-brane
Systems'', JHEP 0505:004, 2005, [arXiv:hep-th/0501190].
\bibitem{newn} N.Seiberg, L.Susskind and N.Toumbas,
``Strings in Background Electric Field, Space/Time Noncommutativity
and A New Noncritical String Theory'', JHEP 0006:021, 2000,
[hep-th/0005040].
\bibitem{ChenSun}B. Chen, M. Li and B. Sun, ``Dbrane near
NS5-branes: with electromagnetic field", JHEP 12(2004)057.
\end{thebibliography}
\end{document}